\documentclass[traditabstract]{aa} 
\usepackage{graphicx}
\usepackage{txfonts}
\usepackage{natbib}
\newcommand{\soft}[1]{{\bf \emph{#1}}}
\usepackage[usenames]{color}

\bibpunct{(}{)}{;}{a}{}{,} 
\begin{document}
   \title{The GalMer database: Galaxy Mergers in the Virtual
Observatory\thanks{The GalMer database is accessible online at
\emph{http://galmer.obspm.fr/}}}

   \author{I. V. Chilingarian\inst{1,2,3}
          \and
          P. Di Matteo\inst{4}
          \and
          F. Combes\inst{1}
          \and
          A.-L. Melchior\inst{1,5}
          \and
          B. Semelin\inst{1,5}
}

   \offprints{Igor Chilingarian \email{igor.chilingarian@obspm.fr}}

   \institute{Observatoire de Paris, LERMA, CNRS UMR~8112, 61 Av. de
l'Observatoire, Paris, 75014, France
         \and
              Sternberg Astronomical Institute, Moscow State University, 13
Universitetski prospect, 119992, Moscow, Russia
         \and
              Centre de donn\'ees astronomiques de Strasbourg, Observatoire
astronomique de Strasbourg, UMR~7550, Universit\'e de Strasbourg / CNRS, 
11 rue de l'Universit\'e, 67000 Strasbourg, France
         \and
             Observatoire de Paris-Meudon, GEPI, CNRS UMR~8111, 5 pl. Jules
Janssen, Meudon, 92195, France
         \and
             Universit\'e Pierre et Marie Curie-Paris 6, 4, Place Jussieu,
F-75\,252 Paris Cedex 05, France
}

   \date{Received February 24, 2010; accepted March 03, 2010; in original
form July 21, 2009}
   \authorrunning{Chilingarian et al.}
   \titlerunning{The GalMer database}

 
  \abstract
{We present the GalMer database, a library of galaxy merger simulations,
made available to users through tools compatible with the Virtual
Observatory (VO) standards adapted specially for this theoretical database.
To investigate the physics of galaxy formation through hierarchical merging,
it is necessary to simulate galaxy interactions varying a large number of
parameters: morphological types, mass ratios, orbital configurations, etc.
On one side, these simulations have to be run in a cosmological context,
able to provide a large number of galaxy pairs, with boundary conditions
given by the large-scale simulations, on the other side the resolution has
to be high enough at galaxy scales, to provide realistic physics. The GalMer
database is a library of thousands simulations of galaxy mergers at moderate
spatial resolution and it is a compromise between the diversity of initial
conditions and the details of underlying physics. We provide all coordinates
and data of simulated particles in FITS binary tables. The main advantages
of the database are VO access interfaces and value-added services which
allow users to compare the results of the simulations directly to
observations: stellar population modelling, dust extinction, spectra,
images, visualisation using dedicated VO tools. The GalMer value-added
services can be used as virtual telescope producing broadband images, 1D
spectra, 3D spectral datacubes, thus making our database oriented towards
the usage by observers. We present several examples of the GalMer database
scientific usage obtained from the analysis of simulations and modelling
their stellar population properties, including: (1) studies of the star
formation efficiency in interactions; (2) creation of old counter-rotating
components; (3) reshaping metallicity profiles in elliptical galaxies; (4)
orbital to internal angular momentum transfer; (5) reproducing observed
colour bimodality of galaxies.}



   \maketitle
%

\section{Introduction}

In the framework of the present cosmological paradigm, mergers and
interactions are among the most important mechanisms governing galaxy
formation and evolution. \citet{SB51} proposed that collisions of
late-type disc galaxies should produce early-type ones. A natural
consequence of this phenomenon is the morphology--density relation
discovered three decades later \citep{Dressler80}: dense regions of the
Universe, where galaxy collisions are supposed to be more frequent, contain
larger fractions of early-type galaxies than sparsely populated areas.

\citet{TT72} were the first to suggest that interactions `tend to bring
\emph{deep} into a galaxy a fairly \emph{sudden} supply of fresh fuel in the
form of interstellar material$\dots$' The gas in the bar formed during the
interaction loses its angular momentum due to the torques \citep{CDG90,BH96}, 
falls onto the galaxy centre possibly inducing strong bursts of star
formation \citep{MH94a,MH96} and creating young compact central stellar
component \citep{MH94b} often observed in present-day early-type galaxies
\citep{Silchenko06,Kuntschner+06}. Intense star formation episodes
accompanied by supernova explosions, enrich the interstellar medium (ISM)
with heavy elements, consumed into the stars hence increasing the observed
metal abundances of the stellar population.

Large scale cosmological $N$-body simulations \citep[e.g.][]{SFW06,OPT08}
often lack the
spatial resolution to trace in detail star formation and morphological
transformation at galaxy scale. Therefore, usually they are complemented by
semi-analytical prescriptions qualitatively accounting for phenomena 
strongly affecting galaxy evolution, such as star formation
\citep[e.g.][]{Blaizot+04,Somerville+08}. However, the
parameters of the semi-analytical models have to be chosen based on more
detailed simulations of galaxy interactions.  High resolution galaxy simulations
\citep[e.g.][]{BDE08} cannot be performed in large statistical numbers.
 A compromise has then to be done between statistics and resolution.
This becomes one of the main motivations for studying large numbers 
of galaxy interactions by means of dedicated
intermediate-resolution numerical simulations.

Merger-induced star formation, as well as morphological transformation,
strongly depends on the mass ratio of the interacting galaxies. Generally,
the intensity of starburst decreases as the merger mass ratio increases
\citep[e.g.][]{Cox+08}. Equal mass mergers of disc galaxies (mass ratios
below 4:1) usually result in early-type elliptical-like remnants
\citep{Toomre77,NB03} while minor mergers (ratios above 10:1) do not destroy
the progenitor's disc preserving its exponential mass distribution although
making it thicker and dynamically hotter \citep{QHF93,WMH96,VW99,BJC05}. A
sequence of repeated minor mergers can form elliptical galaxies, with global
morphological and kinematical properties similar to that observed in real
ellipticals \citep{BJC07}.

Orbital parameters of the interaction and orientation of galaxies also
strongly affect the process of merger, e.g. star formation efficiency on
retrograde orbits is generally higher than for direct encounters
\citep{dMCMS07}.

One needs to explore a large multi-dimensional parameter space (different
initial morphologies related to gas content, orbital configurations, mass
ratios, etc.) by running thousands of simulations in order to fully
understand the astrophysical consequences of galaxy interactions for the
modern picture of galaxy evolution. 

The GalMer project, developed in the framework of the French national
HORIZON\footnote{http://www.projet-horizon.fr/}
collaboration, has the ambitious goal of providing access for the astronomical
community to the results of massive intermediate-resolution numerical
simulations of galaxy interactions in pairs, covering as much as possible
the parameter space of the initial conditions and, thus allowing to study
statistically the star formation enhancements, structural and dynamical
properties of merger remnants. An important aspect is to integrate the data
services into the framework of the International Virtual Observatory in
order to take the full advantage of already developed technologies and data
visualisation and processing tools.

This paper is a technical presentation of: (1) the GalMer TreeSPH
simulations providing all essential details on initial conditions for
galaxies of different masses and prescriptions used to model the processes of
star formation including supernova feedback and metallicity evolution; (2)
the GalMer database, the first VO resource containing
results of TreeSPH simulations; (3) the GalMer value-added services aimed at
facilitating the comparison between simulations and observations such as
modelling the spectrophotometric galaxy properties using an evolutionary
synthesis code.

The paper is organised as follows: in Section~2 we describe initial
conditions of the numerical simulations and orbital parameters of galaxy
interactions; Section~3 contains information on the numerical method,
prescriptions for star formation and metallicity evolution; Section~4
provides the description of the GalMer database structure, its access
interface, and mechanisms of data visualisation; Section~5 presents
value-added services of the GalMer database; in Section~6 we define some
astrophysical applications which can be tackled with our simulations;
Section~7 contains a brief summary.


\section{Initial conditions}
\subsection{Galaxy models: moving along the Hubble sequence}\label{galmod}

We model interactions among galaxies of \emph{different morphologies, from
ellipticals to late-type spirals}. 

The adopted galaxy models consist of a spherical non-rotating dark-matter
halo, which may or may not contain a stellar and a gaseous disc and,
optionally, a central non-rotating bulge. For each galaxy type, the halo and
the optional bulge are modelled as Plummer spheres, with characteristic
masses $M_B$ and $M_H$ and characteristic radii $r_B$ and $r_H$. Our choice
of adopting a core density distribution for the dark halo seems to be more
in accordance with the rotation curves of local spirals and dwarf galaxies,
than the cuspy profiles predicted by Cold Dark Matter simulations
\citep[see][Sect.2.4.3 for a discussion]{DiMatteo+08b}. The stellar and
gaseous discs follow the \citet{MN75} density profile, with masses $M_{*}$ and
$M_{g}$ and vertical and radial scale lengths given, respectively, by
$h_{*}$ and $a_{*}$, and $h_{g}$ and $a_{g}$. We refer the reader to
Appendix \ref{profiles} for the analytical expression of these profiles.  
For the initial models of the disc galaxies, we chose an initial
Toomre parameter for the stellar disc $Q_{\mathrm{star}}=1.2$, and two different
initial ones for the gas component, $Q_{\mathrm{gas}}=0.3$ and $Q_{\mathrm{gas}}=1.2$, in order to
reproduce, at least partially, the variety of parameters found in real
galaxies \citep{MK01,BPBG03,Hitschfeld+09}.

The database contains interacting galaxy pairs of  \emph{different mass
ratios} (1:1, 1:2, 1:10), involving a giant galaxy (gE0 for a giant-like
elliptical, gS0 for a giant-like lenticular, gSa for a giant-like Sa spiral,
gSb for a giant-like Sbc spiral and gSd for a giant-like Sd spiral),
interacting with:
\begin{itemize}
\item either another giant galaxy;
\item or an intermediate-mass galaxy (hereafter iE0, iS0, iSa, iSb and iSd),
having a total mass half that of the giant's mass;
\item or a dwarf galaxy (hereafter dE0, dS0, dSa, dSb and dSd) whose total mass
is ten times smaller that that of the giant galaxy.
\end{itemize}

\begin{figure*}
   \centering
\includegraphics[height=0.19\hsize,angle=270]{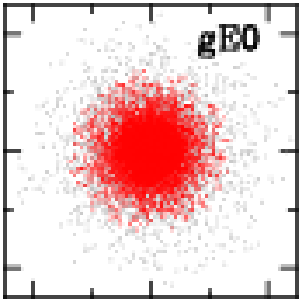}
 \includegraphics[height=0.19\hsize,angle=270]{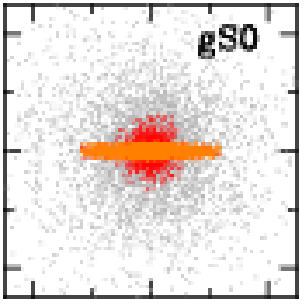}
 \includegraphics[height=0.19\hsize,angle=270]{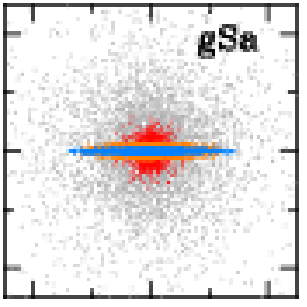}
 \includegraphics[height=0.19\hsize,angle=270]{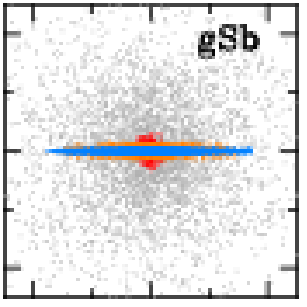}
  \includegraphics[height=0.19\hsize,angle=270]{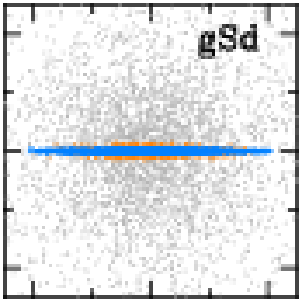}
  \includegraphics[height=0.19\hsize,angle=270]{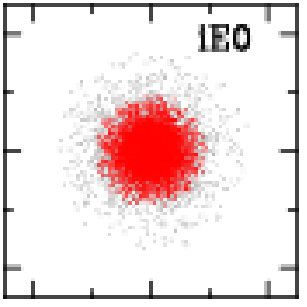}
 \includegraphics[height=0.19\hsize,angle=270]{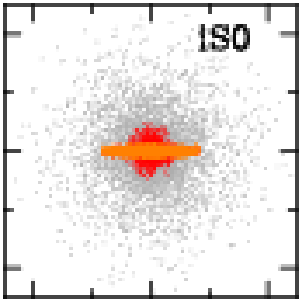}
 \includegraphics[height=0.19\hsize,angle=270]{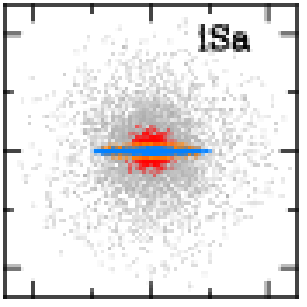}
 \includegraphics[height=0.19\hsize,angle=270]{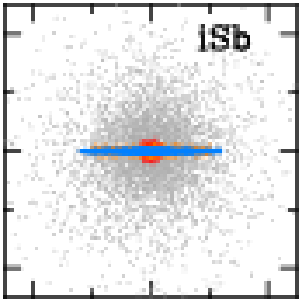}
  \includegraphics[height=0.19\hsize,angle=270]{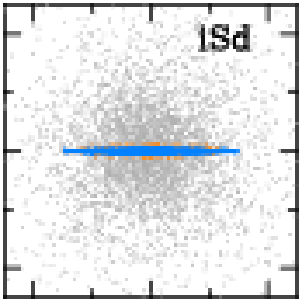}
 \includegraphics[height=0.19\hsize,angle=270]{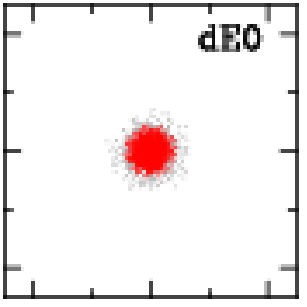}
 \includegraphics[height=0.19\hsize,angle=270]{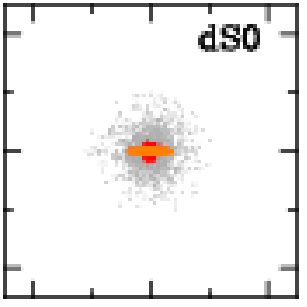}
 \includegraphics[height=0.19\hsize,angle=270]{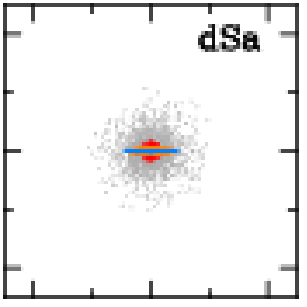}
 \includegraphics[height=0.19\hsize,angle=270]{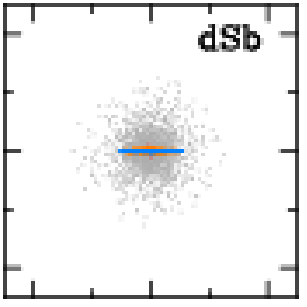}
  \includegraphics[height=0.19\hsize,angle=270]{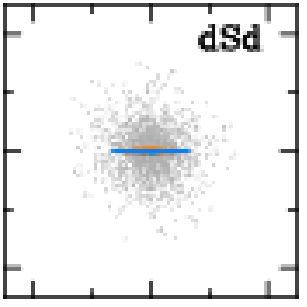}
   \caption{Hubble sequence of the GalMer galaxy models. From left to right, and from the top to the bottom, projection on the x-z plane of giant, intermediate and dwarf galaxies. Different colors correspond to the different galaxy components: dark matter (grey), bulge (red), stellar disc (orange), and gaseous disc (blue). Each frame is 50 kpc $\times$ 50 kpc in size.}.
              \label{allimages}%

    \end{figure*}

The complete Hubble sequence of the galaxy models is given in
Fig.\ref{allimages}. Moving along the Hubble sequence, from gE0 to gSd
galaxies, the mass of the central spheroid varies from $M_B=1.6\times10^{11}
M_{\odot} $ for a gE0 to $M_B=0$ for a gSd, while the gas mass $M_g$, absent
in the case of a gE0 and gS0, increases from $9.2\times 10^9 M_{\odot}$ in a
gSa to $1.7\times10^{10} M_{\odot}$ for a gSd (see Table \ref{galpar1} for a
complete list of all the parameters of giant-like galaxies, and Tables
\ref{galpar2} and \ref{galpar3} for the corresponding set of parameters of
intermediate and dwarf systems). Our modelled Hubble sequence capture the
main properties of local galaxies. We plan in the future to extend this
library to higher redshift, considering larger gas fractions for the disc
galaxies.

\begin{figure}
   \centering
   \includegraphics[height=\hsize,angle=270]{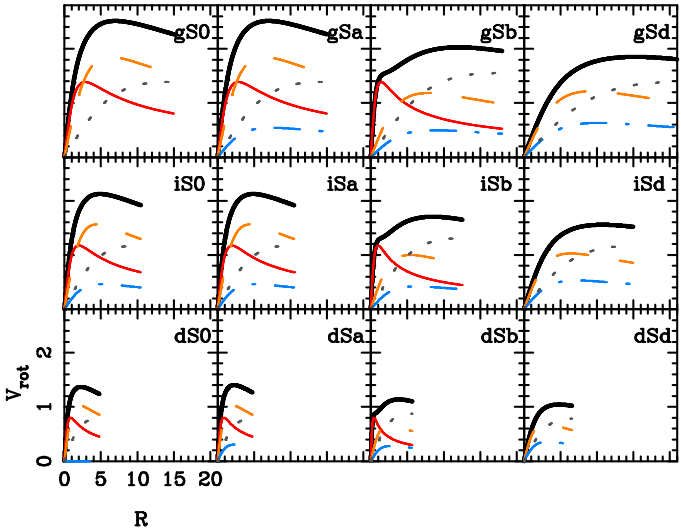}
   \caption{Rotation curves for the different galaxy models. The initial rotation curve is shown (solid thick black curve), together with the contribution  by the bulge component (thin solid red curve), the stellar disc (orange dashed curve), the gaseous disc (dot-dashed blue curve) and the dark matter halo (dotted grey curve). Lengths are in kpc and velocities in units of 100 km/s.}
              \label{allrotcurves}%
    \end{figure}

The initial rotation curves of the modelled disc galaxies are shown in
Fig.\ref{allrotcurves}.


Each galaxy is made up of a total number of particles $N_{\mathrm{TOT}}$ distributed
among gas, stars and dark matter. Gas particles are actually ``hybrid
particles'', characterized by two mass values: one corresponds to the
gravitational mass and stays unchanged during the whole simulation, and the
other is the gas content of the particule, decreasing or increasing
according to the local star formation rate and mass loss (see Sect. 3.1,
Eq.~1). When the gas fraction is below a certain threshold, the particle is
transformed into a star particle, and the remaining gas mass is distributed
over neighbouring gas particles.

For giant-giant interactions, each galaxy is made up of a total number of
particles of $N_{\mathrm{TOT}}$=120000, distributed among gas, stars and dark matter,
depending on the morphology (Table \ref{numbers1}). For giant-intermediate
and giant-dwarf interactions, we increased by a factor 4 the total number of
particles in the giant galaxy.  This allowed us to improve the spatial
resolution of the simulations and, in particular, to maintain a high
enough numerical resolution for the smaller galaxy ($N_{\mathrm{TOT}}$=240000 and
$N_{\mathrm{TOT}}$=48000 for the intermediate and dwarf galaxy, respectively; Tables
\ref{numbers2} and \ref{numbers3}).

To initialize particle velocities, we adopted the method described in
\citet{Hernquist93}.

In order to distinguish the role of interactions from secular evolution, 
  for each galaxy model we provide in the database also its evolution in isolation, for 3 Gyr. A brief description about how the modeled galaxies evolve in isolation is given in Appendix˜\ref{isolated}.

   \begin{table}
      \caption[]{Galaxy parameters for giant galaxies. The bulge and the halo are modelled as Plummer spheres, with characteristic masses $M_B$ and $M_H$ and characteristic radii $r_B$ and $r_H$.  $M_{*}$ and  $M_{g}$ represent the masses of the stellar and gaseous discs whose vertical and radial  scale lengths are given, respectively, by $h_{*}$ and $a_{*}$, and $h_{g}$ and $a_{g}$.}
         \label{galpar1}
         \centering
         \begin{tabular}{lccccc}
            \hline\hline
            & gE0 & gS0 &gSa & gSb & gSd \\
            \hline
            $M_{B}\ [2.3\times 10^9 M_{\odot}]$ & 70 & 10 & 10 & 5 & 0 \\
            $M_{H}\ [2.3\times 10^9 M_{\odot}]$ & 30 & 50 & 50 & 75 & 75\\
            $M_{*}\ [2.3\times 10^9 M_{\odot}]$ & 0 & 40 & 40 & 20 & 25\\
            $M_{g}/M_{*}$ & -- & --& 0.1 & 0.2 & 0.3\\
            & & & & \\
            $r_{B}\ [\mathrm{kpc}]$ & 4 & 2 & 2 & 1& --\\
            $r_{H}\ [\mathrm{kpc}]$ & 7 & 10 & 10 & 12 & 15\\
            $a_{*}\ [\mathrm{kpc}]$ & -- & 4 & 4 & 5 & 6\\
            $h_{*}\ [\mathrm{kpc}]$ & -- & 0.5 & 0.5 & 0.5 & 0.5\\
            $a_{g}\ [\mathrm{kpc}]$ & -- & -- & 5 & 6 & 7\\
            $h_{g}\ [\mathrm{kpc}]$ & -- & -- & 0.2 & 0.2 & 0.2\\
            \hline
         \end{tabular}
   \end{table}

   \begin{table}
      \caption[]{Galaxy parameters for intermediate galaxies. The bulge and the halo are modelled as Plummer spheres, with characteristic masses $M_B$ and $M_H$ and characteristic radii $r_B$ and $r_H$.  $M_{*}$ and  $M_{g}$ represent the masses of the stellar and gaseous discs whose vertical and radial scale lengths are given, respectively, by $h_{*}$ and $a_{*}$, and $h_{g}$ and $a_{g}$.}
         \label{galpar2}
         \centering
         \begin{tabular}{lccccc}
            \hline\hline
            & iE0 & iS0 & iSa & iSb & iSd \\
            \hline
            $M_{B}\ [2.3\times 10^9 M_{\odot}]$ & 35 & 5 & 5 & 2.5 & 0 \\
            $M_{H}\ [2.3\times 10^9 M_{\odot}]$ & 15 & 25 & 25 & 37.5 & 37.5\\
            $M_{*}\ [2.3\times 10^9 M_{\odot}]$ & 0 & 20 & 20 & 10 & 12.5\\
            $M_{g}/M_{*}$ & -- & -- & 0.1 & 0.2 & 0.3\\
            & & & & \\
            $r_{B}\ [\mathrm{kpc}]$ & 2.8 & 1.4 & 1.4 & 0.7& --\\
            $r_{H}\ [\mathrm{kpc}]$ & 5. & 7. & 7. & 8.5 & 10.6\\
            $a_{*}\ [\mathrm{kpc}]$ & -- & 2.8 & 2.8 & 3.5 & 4.2\\
            $h_{*}\ [\mathrm{kpc}]$ & -- & 0.35 & 0.35 & 0.35 & 0.35\\
            $a_{g}\ [\mathrm{kpc}]$ & -- & -- & 3.5 & 4.2 & 5.\\
            $h_{g}\ [\mathrm{kpc}]$ & -- & -- & 0.14 & 0.14 & 0.14\\
            \hline
         \end{tabular}
   \end{table}

   \begin{table}
      \caption[]{Galaxy parameters for dwarf galaxies. The bulge and the halo are modelled as Plummer spheres, with characteristic masses $M_B$ and $M_H$ and characteristic radii $r_B$ and $r_H$.  $M_{*}$ and  $M_{g}$ represent the masses of the stellar and gaseous discs, whose vertical and radial scale lengths are given, respectively, by $h_{*}$ and $a_{*}$, and $h_{g}$ and $a_{g}$.}     
         \label{galpar3}
      \centering
         \begin{tabular}{lccccc}
            \hline\hline
            & dE0 & dS0 & dSa & dSb & dSd \\
            \hline
            $M_{B}\ [2.3\times 10^9 M_{\odot}]$ & 7 & 1 & 1 & 0.5 & 0 \\
            $M_{H}\ [2.3\times 10^9 M_{\odot}]$ & 3 & 5 & 5 & 7.5 & 7.5\\
            $M_{*}\ [2.3\times 10^9 M_{\odot}]$ & 0 & 4 & 4 & 2 & 2.5\\
            $M_{g}/M_{*}$ & -- & -- & 0.1 & 0.2 & 0.3\\
            & & & & \\
            $r_{B}\ [\mathrm{kpc}]$ & 1.3 & 0.6 &  0.6 & 0.3 &  --\\
            $r_{H}\ [\mathrm{kpc}]$ & 2.2 & 3.2 & 3.2 & 3.8 & 4.7\\
            $a_{*}\ [\mathrm{kpc}]$ & -- & 1.3 & 1.3 & 1.6 & 1.9\\
            $h_{*}\ [\mathrm{kpc}]$ & -- & 0.16 & 0.16 & 0.16 & 0.16\\
            $a_{g}\ [\mathrm{kpc}]$ & -- & -- & 1.6 & 1.9 & 2.2\\
            $h_{g}\ [\mathrm{kpc}]$ & -- & -- & 0.06 & 0.06 & 0.06\\
            \hline
         \end{tabular}
   \end{table}

   \begin{table}
      \caption[]{Particle numbers for giant-giant interactions (mass ratio 1:1)}
         \label{numbers1}
     \centering
         \begin{tabular}{lccccc}
            \hline\hline
        & gE0 & gS0 & gSa & gSb & gSd \\
            \hline
        $N_{\mathrm{gas}}$ & -- & -- & 20000 & 40000 & 60000\\
        $N_{\mathrm{star}}$ & 80000 & 80000 & 60000 & 40000 & 20000\\
        $N_{\mathrm{DM}}$ &40000 & 40000 & 40000 & 40000 & 40000\\
            \hline
         \end{tabular}
   \end{table}

   \begin{table}
      \caption[]{Particle numbers for giant-intermediate interactions (mass ratio 1:2)}
         \label{numbers2}
     \centering
         \begin{tabular}{lccccc}
                      \hline\hline
        & gE0 & gS0 & gSa & gSb & gSd \\
            \hline
        $N_{\mathrm{gas}}$ & -- & -- & 80000 & 160000 & 240000\\
        $N_{\mathrm{star}}$ & 320000 & 320000 & 240000 & 160000 & 80000\\
        $N_{\mathrm{DM}}$ &160000 & 160000 & 160000 & 160000 & 160000\\
            \hline\hline
        & iE0 & iS0 & iSa & iSb & iSd \\
            \hline
        $N_{\mathrm{gas}}$ & -- & -- & 40000 & 80000 &120000\\
        $N_{\mathrm{star}}$ & 160000 & 160000 & 120000 & 80000 & \textbf{40000}\\
        $N_{\mathrm{DM}}$ &80000 & 80000 & 80000 & 80000 & 80000\\
            \hline
         \end{tabular}
   \end{table}

   \begin{table}
      \caption[]{Particle numbers for giant-dwarf interactions (mass ratio 1:10)}
         \label{numbers3}
     \centering
         \begin{tabular}{lccccc}
                      \hline\hline
        & gE0 & gS0 & gSa & gSb & gSd \\
            \hline
        $N_{\mathrm{gas}}$ & -- & -- & 80000 & 160000 & 240000\\
        $N_{\mathrm{star}}$ & 320000 & 320000 & 240000 & 160000 & 80000\\
        $N_{\mathrm{DM}}$ &160000 & 160000 & 160000 & 160000 & 160000\\
             \hline\hline
        & dE0 & dS0 & dSa & dSb & dSd \\
            \hline
        $N_{\mathrm{gas}}$ & -- & -- & 8000 & 16000 & 24000\\
        $N_{\mathrm{star}}$ & 32000 & 32000 &  24000 & 16000 & 8000\\
        $N_{\mathrm{DM}}$ &16000 & 16000 & 16000 & 16000 & 16000\\
            \hline
         \end{tabular}
   \end{table}

\subsection{Orbital parameters}\label{orbital}

   \begin{table}
      \caption[]{Orbital parameters for giant-giant interactions}
         \label{orbpos1}
	 \centering
	          \begin{tabular}{cccccc}
       \hline\hline
       orb.id & $r_{\mathrm{ini}}^{\mathrm{a}}$ &  $v_{\mathrm{ini}}^{\mathrm{b}}$  & $L^{\mathrm{c}}$  & $E^{\mathrm{d}}$ & spin$^{\mathrm{e}}$\\
         & kpc & 10$^2$km~s$^{-1}$ & 10$^2$km~s$^{-1}$kpc &  10$^4$km$^2$s$^{-2}$ &\\
            \hline
	    01dir &  100. &2. &56.6 &0.0 & up\\
	    01ret &  100. &2. &56.6 &0.0 & down\\	
	    02dir &  100. &3. &59.3 &2.5 & up\\
	    02ret &  100. &3. &59.3 &2.5 & down\\	
	    03dir &  100. &3.7 &62.0 &5.0 & up\\	
	    03ret &  100. &3.7 &62.0 &5.0 & down\\	
	    04dir &  100. &5.8 &71.5 &15.0 & up\\	
	    04ret &  100. &5.8 &71.5 &15.0 & down\\	
	    05dir &  100. &2. &80.0 &0.0 & up\\	
	    05ret &  100. &2. &80.0 &0.0 & down\\	
	    06dir &  100. &3. &87.6 &2.5 & up\\	
	    06ret &  100. &3. &87.6 &2.5 & down\\	
	    07dir &  100. &3.7 &94.6 &5.0 & up\\	
	    07ret &  100. &3.7 &94.6 &5.0 & down\\	
	    08dir &  100. &5.8 &118.6 &15.0 & up\\	
	    08ret &  100. &5.8 &118.6 &15.0 & down\\	
	    09dir &  100. &2.0 &97.9 &0.0 & up\\	
	    09ret &  100. &2.0 &97.9 &0.0 & down\\	
	    10dir &  100. &3.0 &111.7 &2.5 & up\\	
	    10ret &  100. &3.0 &111.7 &2.5 & down\\	
	    11dir &  100. &3.7 &123.9 &5.0 & up\\	
	    11ret &  100. &3.7 &123.9 &5.0 & down\\	
	    12dir &  100. &5.8 &163.9 &15.0 & up\\	
	    12ret &  100. &5.8 &163.9 &15.0 & down\\
            \hline
         \end{tabular}

\begin{list}{}{}
\item[$^{\mathrm{a}}$] Initial distance between the two galaxies.
\item[$^{\mathrm{b}}$] Absolute value of the initial relative velocity.
\item[$^{\mathrm{c}}$] $L=\mid\bf{r_{\mathrm{ini}}} \times \bf{v_{\mathrm{ini}}}\mid $.
\item[$^{\mathrm{d}}$]  $E={v_{\mathrm{ini}}}^2/2-G(m_1+m_2)/r_{\mathrm{ini}}$, with $m_1=m_2=2.3
\times10^{11}M_{\odot}$.
\item[$^{\mathrm{e}}$]   Orbital spin, if the z-component is parallel (up) or anti-parallel (down) to the z-axis
\end{list}

   \end{table}

   \begin{table}
      \caption[]{Orbital parameters for giant-intermediate interactions}
         \label{orbpos2}
	 \centering
         \begin{tabular}{cccccc}
            \hline\hline
             orb.id & $r_{\mathrm{ini}}^{\mathrm{a}}$ &  $v_{\mathrm{ini}}^{\mathrm{b}}$  & $L^{\mathrm{c}}$  & $E^{\mathrm{d}}$ & spin$^{\mathrm{e}}$\\
         & kpc & 10$^2$km~s$^{-1}$ & 10$^2$km~s$^{-1}$kpc &  10$^4$km$^2$s$^{-2}$ &\\
            \hline
	    01dir &  100. &1.73 &48.0 &0.0 & up\\
	    01ret &  100. &1.73 &48.0 &0.0 & down\\	
	    02dir &  100. &1.79 &49.0 &0.1 & up\\
	    02ret &  100. &1.79 &49.0 &0.1 & down\\	
	    03dir &  100. &1.73 &60.0 &0.0 & up\\	
	    03ret &  100. &1.73 &60.0 &0.0 & down\\	
	    04dir &  100. &1.79 &61.0 &0.1 & up\\	
	    04ret &  100. &1.79 &61.0 &0.1 & down\\	
	    05dir &  100. &1.73 &69.3 &0.0 & up\\	
	    05ret &  100. &1.73 &69.3 &0.0 & down\\	
	    06dir &  100. &1.79 &69.7 &0.1 & up\\	
	    06ret &  100. &1.79 &69.7 &0.1 & down\\	
            \hline
         \end{tabular}
\begin{list}{}{}
\item[$^{\mathrm{a}}$] Initial distance between the two galaxies.
\item[$^{\mathrm{b}}$] Absolute value of the initial relative velocity.
\item[$^{\mathrm{c}}$] $L=\mid\bf{r_{\mathrm{ini}}} \times \bf{v_{\mathrm{ini}}}\mid $.
\item[$^{\mathrm{d}}$]  $E={v_{\mathrm{ini}}}^2/2-G(m_1+m_2)/r_{\mathrm{ini}}$, with $m_1=2.3
\times10^{11}M_{\odot}$ and $m_2=1.15\times10^{10}M_{\odot}$.
\item[$^{\mathrm{e}}$]   Orbital spin, if the z-component is parallel (up) or anti-parallel (down) to the z-axis
\end{list}
   \end{table}

   \begin{table}
      \caption[]{Orbital parameters for giant-dwarf interactions}
         \label{orbpos3}
	 \centering
         \begin{tabular}{cccccc}
       \hline\hline
       orb.id & $r_{\mathrm{ini}}^{\mathrm{a}}$ &  $v_{\mathrm{ini}}^{\mathrm{b}}$  & $L^{\mathrm{c}}$  & $E^{\mathrm{d}}$ & spin$^{\mathrm{e}}$\\
         & kpc & 10$^2$km~s$^{-1}$ & 10$^2$km~s$^{-1}$kpc &  10$^4$km$^2$s$^{-2}$ &\\
       \hline
       01dir & 100.& 1.48&29.66&0.& up\\
       01ret & 100.& 1.48&29.66&0.& down\\
       02dir & 100.& 1.52&29.69&0.05& up\\
       02ret & 100.& 1.52&29.69&0.05& down\\
       03dir & 100.& 1.55&29.72&0.1& up\\
       03ret & 100.& 1.55&29.72&0.1& down\\
       04dir & 100.& 1.48&36.33&0.0& up\\
       04ret & 100.& 1.48&36.33&0.0& down\\
       05dir & 100.& 1.52&36.38&0.05& up\\
       05ret & 100.& 1.52&36.38&0.05& down\\
       06dir & 100.& 1.55&36.43&0.1& up\\
       06ret & 100.& 1.55&36.43&0.1& down\\
       \hline  
       \hline
         \end{tabular}

\begin{list}{}{}
\item[$^{\mathrm{a}}$] Initial distance between the two galaxies.
\item[$^{\mathrm{b}}$] Absolute value of the initial relative velocity.
\item[$^{\mathrm{c}}$] $L=\mid\bf{r_{\mathrm{ini}}} \times \bf{v_{\mathrm{ini}}}\mid $.
\item[$^{\mathrm{d}}$]  $E={v_{\mathrm{ini}}}^2/2-G(m_1+m_2)/r_{\mathrm{ini}}$, with $m_1=2.3
\times10^{11}M_{\odot}$ and $m_2=2.3\times10^{10}M_{\odot}$.
\item[$^{\mathrm{e}}$]   Orbital spin, if the z-component is parallel (up) or anti-parallel (down) to the z-axis
\end{list}
   \end{table}

\begin{figure}
   \centering
   \includegraphics[width=8cm,angle=0]{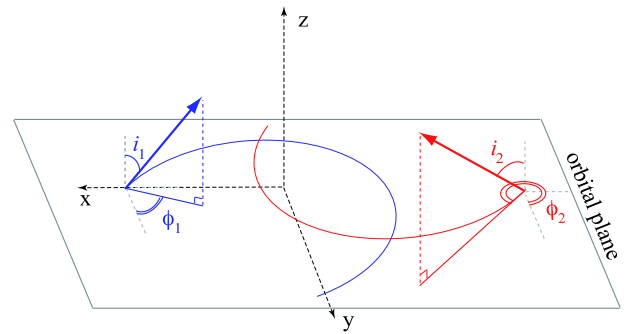}
   \caption{Adopted orbital geometry for the simulations. We set up the collision
in such a way that the orbital angular momentum is parallel to the z-axis and that the centers of the two galaxies initially are on the x-axis. The galaxy spins are represented by the blue and red arrows, respectively. They are specified in terms of the spherical coordinates
($i_1$,$\Phi_1$) and ($i_2$,$\Phi_2$). See Table \ref{angles} for their initial values.}
              \label{fangles}%
    \end{figure}

For each pair of interacting galaxies, we performed several simulations,
varying the galaxies' orbital initial conditions (initial orbital energy $E$
and angular momentum $L$) and taking into account both direct and retrograde
orbits (Tables \ref{orbpos1}, \ref{orbpos2} and \ref{orbpos3}).  For each
interacting pair, we kept the disc (when present) of one of the two galaxies
in the orbital plane ($i_1=0^\circ$), and varied the inclination $i_2$ of the
companion disc, considering: $i_2=0^\circ$, $i_2=45^\circ$, $i_2=75^\circ$, and
$i_2=90^\circ$.  The clustering of the angles toward $i_2=90^\circ$, for an
uneven sampling, is logical from a pure geometrical point of view,
considering that the probability of the spin $i_2$ of the second galaxy to
be oriented between 0 and $i_2$ is proportional to $1-\cos(i_2)$: this means,
for example, that an orientation $i_2$ between 45$^\circ$ and 90$^\circ$ has
probability 2.3 times higher than an orientation $i_2$ between 0$^\circ$ and
45$^\circ$. However, the spins alignment may not be totally uncorrelated, as
recently shown by \citet{Jimenez+10}, but a distribution function, to our
knowledge, is still lacking. For giant-dwarf interactions, we also consider
a more generic case, with $i_1=33^\circ$, and $i_2=130^\circ$ (see Fig.
\ref{fangles} for a sketch of the initial orbital geometry and Table
\ref{angles} for the orientation of the galaxy spins).

   \begin{table}
         \caption[]{Orientation of the galaxy spins, for giant-giant, giant-intermediate, and giant-dwarf simulations.}
         \label{angles}
     \centering
         \begin{tabular}{lcc}
            \hline\hline
	    & giant-giant & giant-dwarf\\
	    & giant-intermediate & \\
	    	    & giant-dwarf & \\
	    \hline
	    $i_1$ & 0$^\circ$ & 33$^\circ$\\
	    $\Phi_1$ & 0$^\circ$ & 0$^\circ$\\
	    $i_2$ & 0$^\circ$,45$^\circ$,75$^\circ$,90$^\circ$ & 130$^\circ$\\
	    $\Phi_2$ & 0$^\circ$ & 0$^\circ$\\
            \hline
         \end{tabular}
   \end{table}

\section{Numerical method}\label{Numericalmethods}

To model galaxy evolution, we employed a Tree-SPH code, in which
gravitational forces are calculated using a hierarchical tree method
\citep{BH86} and gas evolution is followed by means of smoothed particle
hydrodynamics \citep[SPH,][]{Lucy77,GM82}. Gravitational forces are calculated using
a tolerance parameter $\theta=0.7$ and include terms up to the quadrupole
order in the multipole expansion. A Plummer potential is used to soften
gravitational forces, with same softening lengths for all
particle types. We assume a softening length $\epsilon=280\ \mathrm{pc}$ for
giant-giant interactions, and $\epsilon=200\ \mathrm{pc}$ for the
giant-intermediate and giant-dwarf runs.

The code (evaluation of the gravitational forces, implementation of the SPH
technique and star formation modelling) was described in \citet{SC02},
where a standard validation test for this type of codes (the collapse of an
initially static, isothermal sphere of self-gravitating gas, see
\citet{Evrard88} and also \citet{HK89, Thacker+00}, \citealp{SYW01}) was
also presented. In our present study, we adopted an isothermal gas
phase for the GalMer runs.  Other tests are presented in
\citet{DiMatteo+08b}.

Smoothed Particle Hydrodynamics (SPH) is a Lagrangian technique in which the
gas is partitioned into fluid elements represented by particles, which obey
equations of motion similar to the collisionless component, but with
additional terms describing pressure gradients, viscous forces and radiative
effects in gas. To capture shocks a conventional form of the artificial
viscosity is used, with parameters $\alpha=0.5$ and $\beta=1.0$
\citep{HK89}.  To describe different spatial dynamical ranges, SPH particles
have individual smoothing lengths $h_i$, calculated in such a way that a
constant number of neighbours lies within $2h_i$. The giant-giant
simulations were performed using a number of neighbours $N_s\sim 15$;
for the giant-intermediate and giant-dwarf interactions $N_s\sim 50$.  The
gas is modelled as isothermal, with a temperature $T_{\mathrm{gas}}=10^4 K$. Because
of the short cooling time of disc gas, fluctuations in the gas temperature
are quickly radiated away, so that simulations employing an isothermal
equation of state differ little from more realistic ones \citep{MH96,NJB06}.

The equations of motion are integrated using a leapfrog algorithm with a
fixed time step of $\Delta t=5\times10^5 \mathrm{yr}$.

\subsection{Star Formation and continuous stellar mass loss}\label{sfevol}

Numerous prescriptions and techniques exist
\citep{Katz92,SM94,Springel00,SH03,CJPS06} for modelling star formation
rate and feedback in numerical simulations.

As in \citet{MH94b}, we parametrized the star formation efficiency for a
SPH particle as

\begin{equation}\label{loc}
\frac{\dot{M}_{\mathrm{gas}}}{M_{\mathrm{gas}}}=C\times {\rho_{\mathrm{gas}}}^{1/2}
\end{equation} 

\noindent
with the constant $C=0.3\, \mathrm{pc}^{3/2}{M_{\odot}}^{-1/2}\mathrm{Gyr}^{-1}$  such that the isolated giant disc galaxies form
stars at an average rate of between $ 1$ and $2.5 M_{\odot} yr^{-1}$. 

The choice of the parametrization in Eq.~\ref{loc} is consistent with the
observational evidence that on global scales the SFR in disc galaxies is
well represented by a Schmidt law of the form
$\Sigma_{\mathrm{SFR}}=A{\Sigma_{\mathrm{gas}}}^N$,  $\Sigma_{\mathrm{gas}}$ and $\Sigma_{\mathrm{SFR}}$ being
disc-averaged surface densities, with the best fitting slope $N$ about 1.4
\citetext{see \citealp{Kennicutt98}, but also \citealp{WB02,BPBG03,GS04}}.
This relation seems to apply, with a similar slope, also to local scales, as
shown in \citet{Kennicutt+05} for Messier~51. As we checked, the
prescriptions we adopted well reproduces the Schmidt law in the
$\Sigma_{\mathrm{gas}}-\Sigma_{\mathrm{SFR}}$ plane, both for isolated and interacting
galaxies \citep{dMCMS07}.

Once the SFR recipe is defined, we apply it to SPH particles, using
the hybrid method described in \citet{MH94b}. In this method, each
initial gas particle is in fact hybrid containing a gas fraction and a star
fraction. Its gravitational mass $M_i$ remains constant, 
but the gas content of the
particle, $M_{i,\mathrm{gas}}$, changes over time according to
Eq.~\ref{loc}. Gravitational forces are always evaluated using the
gravitational mass $M_i$, while hydrodynamical quantities, in turn,
use the time-varying $M_{i,\mathrm{gas}}$.  If the gas fraction present in the
hybrid particles drops below $5\%$ of the initial gas content, the
hybrid particle is totally converted into a star-like particle and the
small amount of gas material still present is spread over the
gas fraction of the neighbours. 

The effects of star formation on the surrounding ISM were
implemented using the technique described in \citet{MH94b}. For each
star-forming hybrid particle, we evaluate the number of stars formed with
masses $> 8 M_{\odot}$ (a \citet{MS79} stellar initial mass function (IMF)
is adopted), and we assume that they instantaneously become supernovae,
leaving behind remnants of $1.4 M_{\odot}$ and releasing their mass to the
surrounding ISM. The released mass also enriches the metallicity of the
surrounding gas. This is done assuming a yield $y=M_{ret}/M_{*}$=0.02, where
$M_{ret}$ is the total mass of all reprocessed metals and $M_{*}$ the total
mass in stars. For each gas particle, mass and metals return is applied to
the $i$-th neighbour gas particle, using a weight $w_i$ based on the
smoothing kernel.

The energy injection in the ISM from SNe explosions is accounted
assuming that only a fraction $\epsilon_{\mathrm{kin}}$ of $E_{SN}=10^{51}
\mathrm{erg}$ goes into kinetic energy, by applying a radial kick to
velocities of neighbour gas particles; thus, for each SNe explosion, the
$i$-th neighbouring gas particle receives a velocity impulse directed
radially away from the ``donor''
\begin{equation}\label{dv}
 \Delta v_{i}=\left(\frac{2 \ w_i \ \epsilon_{\mathrm{kin}} \ E_{SN}}{M_i}\right)^{1/2}, 
\end{equation}
$w_i$ being the weighting based on the smoothing kernel and
$M_i$ the mass of the receiver.

The value of  $\epsilon_{\mathrm{kin}}$ is chosen such that the total amount of
kinetic energy received by a gas particle, due to the contribution from all
its neighbours, is $\leq 1$~km~s$^{-1}$ to prevent a rapid
growth of the vertical thickness of the gaseous disc.

Together with star formation, we also take into account  
the competing process of stellar mass-loss:   at each time step, an amount
\begin{equation}
M_{i,s}(t)=\frac{\left(M_i-M_{i,\mathrm{gas}}(t)\right)\Delta t\ c_0}{t-t_{\mathrm{birth}}+T_0}
\end{equation}
of the stellar mass formed in the hybrid particle is assumed lost by
evolutionary effects, going to enrich the gas content $M_{i,\mathrm{gas}}$ of
the particle. In the formula above, $t_{\mathrm{birth}}$ represents the birth
time of the population, $T_0=4.97$~Myr and $c_0=5.47\times10^{-2}$
\citep[see][for details]{JCPv01}.

\subsection{Metallicity evolution}\label{metevol}

The metallicity content of the modelled giant galaxies is initially
distributed according to a radial profile
\begin{equation}\label{gradient}
z_{m}(\textbf{R})= z_{0} \times 10^{-0.07*R}
\end{equation}
where $R$ is the particle distance from the galaxy center and
$z_{0}=3\times z_{\odot}$ \citep{KBG03,Magrini+07,Lemasle+08}. 
For intermediate and dwarf galaxies, we generalised this 
formula, taken account a dependency on the galaxy mass and on its 
half-mass radius \citep{Tremonti+04,Lee+06}:
\begin{eqnarray}\label{gradient2}
z_{m}(R) =  \sqrt{M_{\mathrm{gal}}/M_{\mathrm{giant}}}\, z_{0} \times 10^{-0.07*4.85*R/r_{50}} \nonumber \\
=\sqrt{M_{\mathrm{gal}}/M_{\mathrm{giant}}} z_{0} \times 10^{-0.34*R/r_{50}}
\end{eqnarray}
$M_{\mathrm{gal}}$ being the mass of the intermediate (or dwarf galaxy),  $r_{50}$
its half-mass radius, $M_{\mathrm{giant}}$ the mass of the giant galaxy, and
$r_{50}=4.85$ kpc the average half-mass radius of giant galaxies in the
sample.

The metallicity of the old stellar component is kept unchanged during the
simulations, so that only remixing and dynamical effects can reshape the
initial gradient of the old stellar population. In turn, the metallicity of
the gas component and of the new stellar population (that formed during the
simulation) changes with time, due to the release of metals from SNe
explosions, as star formation proceeds. In more detail, hybrid particles
are characterised by two metallicity values: $z_{m}$ and $z_{\mathrm{new}}$.
The first value corresponds to the metallicity of the gas mass of the particle, 
while the second one provides the metallicity of the stellar mass contained in the hybrid
particle. As for old stars, initially their metallicity $z_{m}$ is
distributed into the galaxy disc accordingly to Eq.~\ref{gradient}, with the
central regions more metal-rich than the outer disc. The metallicity
$z_{\mathrm{new}}$ of a new stellar component, is set  equal to that of the gas
in which it forms, while the reprocessed metals enrich the
surrounding gas, according to the yield described before.

\section{The GalMer Database} 

All the results of the GalMer simulations are accessible online using
Virtual Observatory technologies. The three essential blocks for providing
online access to the data are: (1) storing the data; (2) storing and
querying the data description, i.e. metadata; (3) mechanisms for accessing
and visualising the data. Besides, we provide services to perform online
data analysis, which are described in detail in the next section.

\subsection{Data format and storage}

Aimed at interoperability and performance, we chose the FITS Binary
Table format \citep{CTP95,Hanisch+01} to store the simulation results. The
FITS format is handled by a variety of tools widely used by the astronomical
community. A FITS binary table can be easily incorporated into the VOTable
format \citep{Ochsenbein+04} used by new generation astronomical software
tools introduced by the Virtual Observatory.

Every galaxy interaction (``GalMer experiment'' hereafter) includes 50 to 70
snapshots with a 50~Myr time interval containing data for individual
particles traced by the simulations, thus following the evolution of an
interacting galaxy pair for 2.5--3.5~Gyr. We store each snapshot as an
individual file. The following properties are provided for every particle:
Cartesian coordinates ($X$, $Y$, $Z$), three-dimensional velocity vector
($v_X$, $v_Y$, $v_Z$), total mass ($M$), particle type (\emph{hybrid},
\emph{star} or \emph{dark matter}), identification of a galaxy which a given
particle belonged to in the initial step of the simulation.  Besides, we
provide metallicities and $t_{\mathrm{birth}}$, i.e. the average birth time of the
stellar material in a given particle; for stellar particles, they are kept
fixed through the simulations, since these particles do not evolve. For
\emph{hybrid} particles we provide mean $t_{\mathrm{birth}}$ and metallicities $z_m$
instead, as well as gas masses, metallicities $z_{\mathrm{new}}$ of stars
formed during the previous timestep (i.e. current gas metallicity).

All the information related to the input parameters of a given GalMer
experiment: morphological types of galaxies, orbital configuration, units of
masses, coordinates, and velocities, as well as the epoch of a current
snapshot are provided in FITS headers of snapshot files, making them
self-consistent and available for further stand-alone analysis without need
to connect to the GalMer database.

A typical snapshot containing 240000 particles has a size of 12~MBytes,
resulting in 0.6$-$0.9~GBytes per GalMer experiment. The present data
release contains simulations of about a thousand giant--giant interactions
having a total volume of $\sim$0.9~TBytes. New simulations will be
ingested into the database and put online as soon as they have been
completed.

\subsection{Simulation metadata}

The metadata are computed for every individual snapshot at the time of the
database update. Since the data are archived and do not change in time,
neither the metadata do, the access to the database is read-only, unless new
simulation results are ingested into it.

The metadata of GalMer simulations conform to the current version of the
SimDB data model (Lemson et al. in prep.) being presently developed by the
International Virtual Observatory Alliance (IVOA). SimDB is supposed to
provide a complete self-sufficient description of $N$-body simulations results
using object-oriented approach and is designed using Unified Modeling
Language (UML). For practical usage, e.g. for constructing a database
containing the numerical simulation metadata, a UML data model has to be
serialized. We partly serialize the SimDB data model into a relational
database schema keeping another part (modified Characterisation class of
SimDB) serialized into native XML.

We use the open-source object-relational database
management system (DBMS) \soft{PostgreSQL}\footnote{http://www.postgresql.org/} to implement the advanced
metadata querying mechanisms described in detail in \citet{ZSBC07}. We 
modified the original SimDB data model by replacing its Characterisation
object with the full IVOA Characterisation Data Model (CharDM) metadata
\citep{Louys+08}. CharDM is a way to say where, how extended and in which
way the observational or simulated dataset can be described in a
multidimensional parameter space. Our metadata querying approach allows us
to use additional \emph{WHERE} clauses in standard SQL queries expressed as
XPath statements to constrain particular CharDM elements without need to
serialize rather complex CharDM structure into a relational database schema
and, therefore, prevents additional complications of SQL queries.

The relational DBMS stores metadata including links to FITS files containing
all coordinates and data of
individual simulation particles. Therefore, to execute operations involving
particles, such as cutouts or statistics, it is necessary to fetch the
actual datasets. However, all global properties of simulations, such as
total masses of gas and stars are available inside the DBMS making possible
to extract a global star formation history of a given GalMer experiment at
the database level. This function is implemented as a stored procedure
inside the SQL database and is accessible from the web-interface.

All actual metadata queries are executed by the server-side of the
database web-interface using the parameters visualised in web-pages,
therefore user does not need to type in SQL queries inside web-based forms.

\subsection{Data access and visualisation}

\begin{figure*}
\includegraphics[width=\hsize]{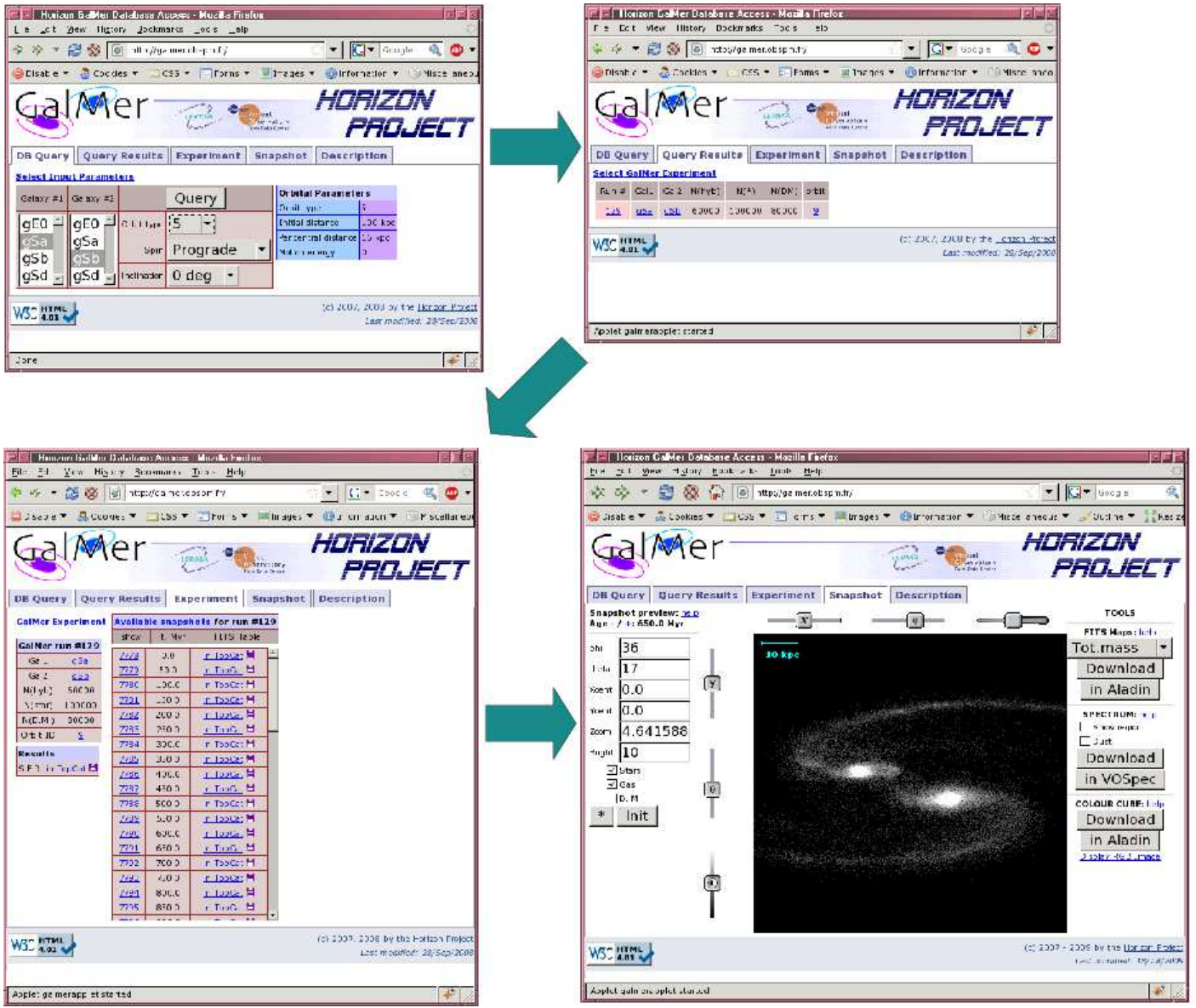}
\caption{Interactive data access using the GalMer web-interface: (1)
selecting galaxy morphologies and orbital types; (2) choosing one GalMer
experiment from the list of those matching the selection criteria; (3)
selecting a snapshot; (4) previewing the snapshot and accessing data
analysis tools for it. \label{figQuery}}
\end{figure*}

The interactive access to the data is provided through the WWW interface at
\emph{http://galmer.obspm.fr/}. It is based on a \emph{de facto} standard
asynchronous JavaScript and XML (\soft{AJAX}) technology and, thus, supports
most widely-used modern Internet-browsers: \soft{Mozilla Firefox}~ver.$>$1.5,
\soft{Microsoft Internet Explorer}~ver.$>$5, \soft{Apple Safari}~ver.$>$3.
Individual snapshots for every simulation can be directly accessed in a
batch mode as well.

The web-site provides a database query interface for accessing simulations
for given galaxy morphological types and orbital parameters of interactions. 
Being a part of SimDB metadata, this information is stored in the database,
and is retrieved dynamically and displayed in a pop-up window, once user has
selected particular elements in the query form. In order to use all
available features of the GalMer database web-interface, the following
software components have to be installed/enabled on the user's computer: (1)
JavaScript support in a web browser; (2) \soft{SUN
Java}\footnote{http://www.java.com/} including the Java Applet browser
plug-in and support for the Java WebStart funcionality which are normally
configured automatically during the installation of \soft{SUN Java}.

The interactive data access includes several steps (see Fig.~\ref{figQuery}).
At first, galaxy morphological types and orbital configurations of an
interaction are chosen using the ``DB Query'' tab. Then, the user is asked
to select one GalMer experiment in the``Query Results'' tab from a list of
those matching the selection criteria provided at the first step. Once an
experiment has been selected, the user can download or visualise its
integrated star formation history and/or download individual snapshots
provided in the ``Experiment'' tab. Then, it is possible to preview the
contents of a given snapshot and get access to the value-added data analysis
tools for it using the ``Shapshot'' tab.

In the ``Snapshot'' tab we provide a powerful \soft{AJAX}-based preview
mechanism for interactive data manipulation directly from within the
web-browser (see bottom right panel in Fig.~\ref{figQuery}). The main purpose
of the snapshot visualisation and manipulation is an interactive choice of
the projection and scaling parameters for the generation of projected maps
and synthetic images and/or spectra as described in the next section.

Although the interactive data visualisation capabilities available
inside the web-browser remain limited, this limitation can be overpassed by
using dedicated software tools if the data are sent to them directly from
the database web-interface. 

\begin{figure}
\includegraphics[width=\hsize]{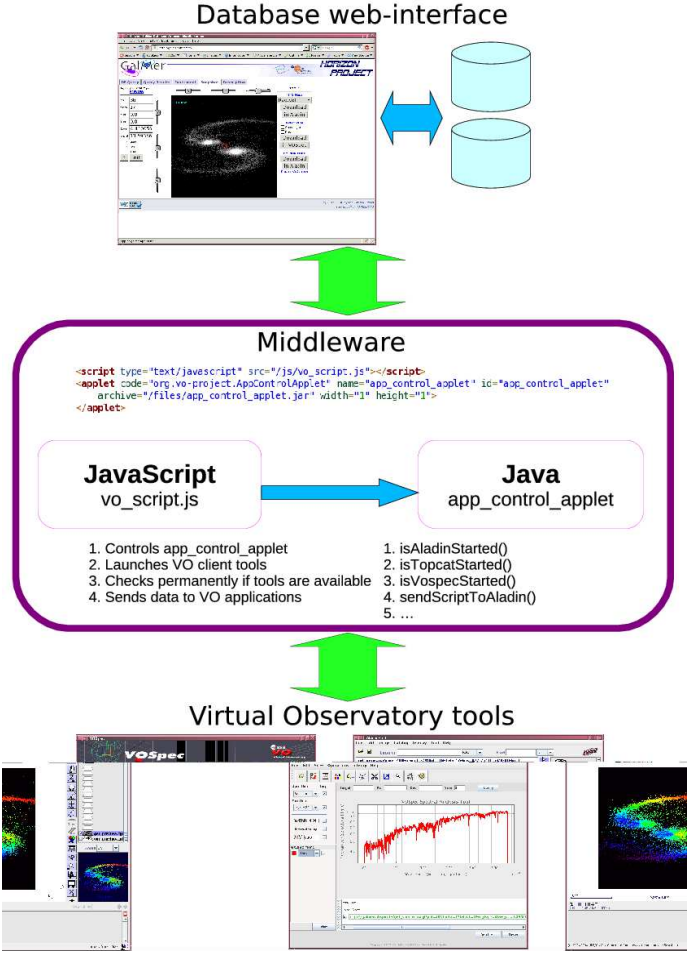}
\caption{The middleware connecting the GalMer database web-interface with
the VO tools dedicated for advanced data manipulation
at top, is illustrated a projection with \soft{mergerapi}, and at bottom, a VOSPEC
spectrum, and a map in Aladin.
\label{midware}}
\end{figure}

We implemented the interaction between the GalMer database
web-interface and existing Virtual Observatory tools dedicated for dealing
with tables
(\soft{topcat}\footnote{http://www.star.bris.ac.uk/$\sim$mbt/topcat/},
\citealp{Taylor05}), images (\soft{CDS
Aladin}\footnote{http://aladin.u-strasbg.fr/}, \citealp{Bonnarel+00}), and
spectra (\soft{ESA VOSpec}\footnote{http://esavo.esa.int/vospec/},
\citealp{OBSL08}). The mechanism of interaction was proposed by \citet{CZ08}
and initially implemented for the observational data archive ASPID-SR
\citep{ASPIDSR}. It is based on the middleware \citep{ZC08} represented by
the two components (Fig.~\ref{midware}) interacting using Sun Java
LiveConnect: (1) a part handling user's actions inside the web-browser
implemented in JavaScript and (2) a Java applet for sending data to the VO
applications, implementing a VO application messaging protocol.
Present version of the middleware implements the \soft{PLASTIC} (\soft{
PLatform for AStronomical Tool InterConnection}, \citealp{Boch+06})
protocol. Support for the more advanced \soft{SAMP} (Simple Application
Messaging Protocol, \citealp{Taylor+09}) will be added in the next release
of the database.

\section{Data Analysis Services}

We developed an application programming interface (API) library to
access and analyse the results of GalMer simulations. The \soft{mergerapi}
library is implemented in \soft{ANSI C} and needs a minimal set of
prerequisites to compile: the \soft{cfitsio} FITS API
library\footnote{http://heasarc.gsfc.nasa.gov/docs/software/fitsio/fitsio.html}
to access the simulation data and the \soft{GD}
library\footnote{http://www.libgd.org/} to generate \emph{gif} and
\emph{png} output images directly viewable inside a web-browser. The \soft{
mergerapi} library is installed on the server side and is used by several
on-the-fly data analysis services described hereafter.

\subsection{Projected Maps} 

For a given snapshot we provide a set of services for the on-the-fly
server-side computation of projected maps of various quantities traced by
the simulations. These tools are accessible at the ``Snapshot'' tab at
the GalMer web-site. 

The following quantities can be computed for any combination of the three
particle types: \emph{surface density}, \emph{line-of-sight radial velocity}
($v_r$), \emph{line-of-sight velocity dispersion} ($\sigma$).  \emph{Gas
mass}, \emph{stellar mass}, \emph{gas metallicity}, and \emph{metallicity of
new stars} apply to \emph{hybrid} particles only. \emph{Stellar metallicity}
provides a metallicity of underlying (``old'') stellar population, therefore
it applies to ``star'' particles and remains constant in time for a given
particle. Preview surface density maps are shown directly in a web-browser,
while for other quantities (velocities, metallicities, etc.) the maps are
available for download as FITS images or can be visualised with \soft{CDS
Aladin}.

In order to generate maps on-the-fly, we developed an efficient
computational algorithm. All the maps are computed for a parallel
projection onto a plane (i.e. assuming an observer at the infinite distance)
at the same time allowing to specify a spatial scale (i.e. to give an
effective distance from an observer to the barycentre $r$). Therefore, the
viewport is uniquely defined by the two quantities: azimuthal ($\varphi$)
and polar ($\theta$) angles. To simplify the comparison of computed maps
with observations and mosaicing of maps with different spatial sampling
(e.g. zoom-in on a given regions of a merger remnant with the overall image)
we use the FITS WCS convention \citep{GC02} assuming a tangential projection
and assigning right ascension and declination of zero to the projected
barycentre position. In this case, the coordinates of a particle on the
projected plane are computed as:

\begin{eqnarray}\label{xyproj}
x_{\mathrm{proj}} = - X \sin{\varphi} + Y \cos{\varphi} \nonumber\\
\eta = 206265 & x_{\mathrm{proj}} / r \nonumber\\
y_{\mathrm{proj}} = - X \cos{\varphi} \sin{\theta} - Y \sin{\varphi} \sin{\theta} + Z
\cos{\theta}\nonumber\\
\xi = 206265 & y_{\mathrm{proj}} / r
\end{eqnarray}

Here $\eta$ and $\xi$ define the tangential coordinates on the sky in
arcsec which are used to compute synthetic images comparable directly to
observations as demonstrated in Section~6.2.

The radial velocity and the position on the line of sight are given by:
\begin{eqnarray}\label{vrproj}
v_{\mathrm{r}} = v_X \cos{\varphi} \cos{\theta} + v_Y \sin{\varphi} \cos{\theta} +
v_Z \sin{\theta}\\
z_{\mathrm{r}} = X \cos{\varphi} \cos{\theta} + Y \sin{\varphi} \cos{\theta} + 
Z \sin{\theta}
\end{eqnarray}

Then,
\begin{itemize}
\item A grid corresponding to the desired map size and sampling is created.
\item In a single loop over all particles, we compute, which bin each
particle will contribute to, and take into account this contribution.
\item In a second loop over the bins (pixels) we compute (if necessary)
the actual values of a physical parameter. 
\end{itemize}

\begin{figure}
\includegraphics[width=0.49\hsize]{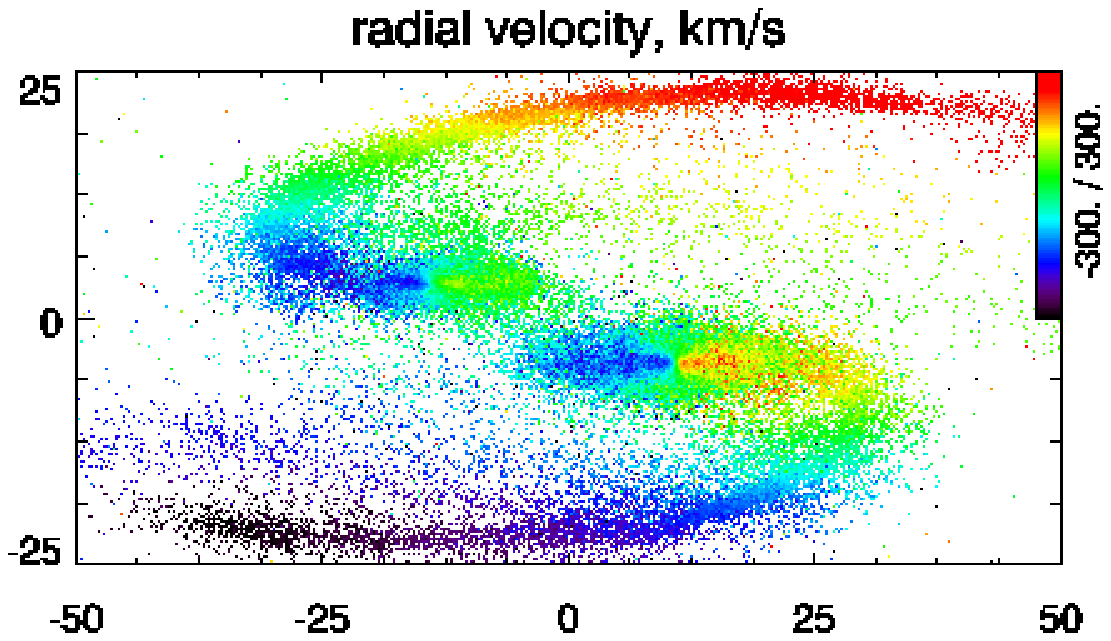}
\includegraphics[width=0.49\hsize]{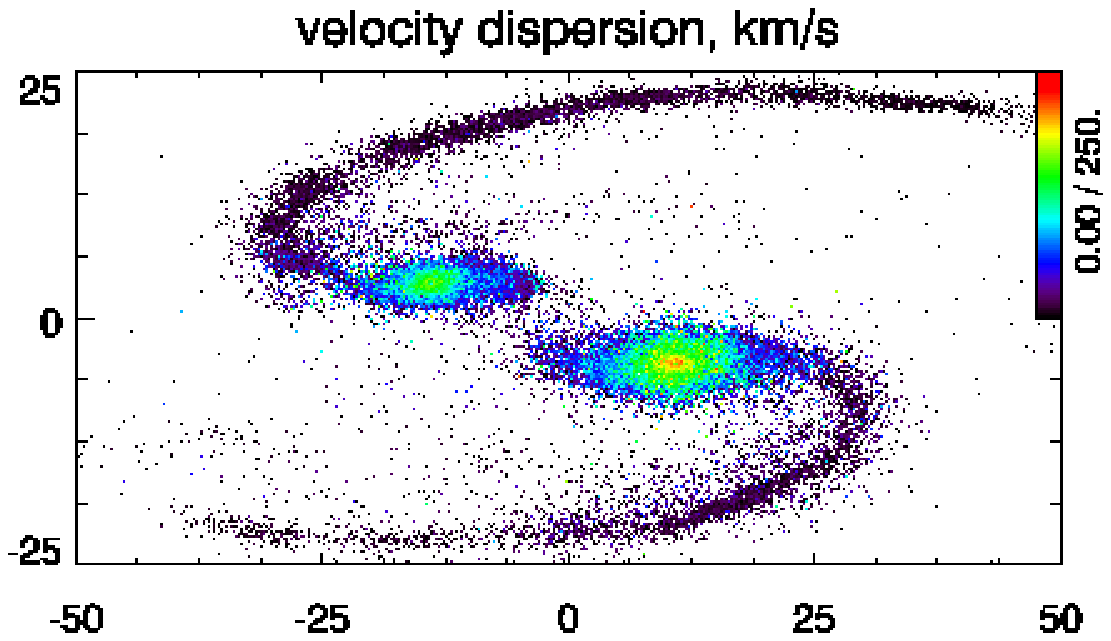}\\
\includegraphics[width=0.49\hsize]{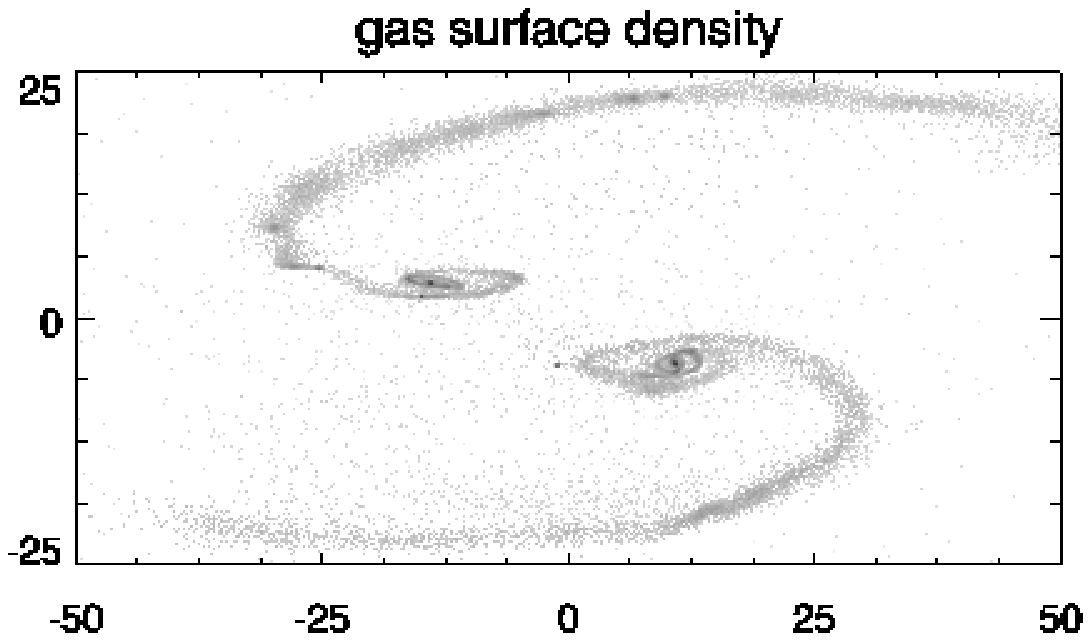}
\includegraphics[width=0.49\hsize]{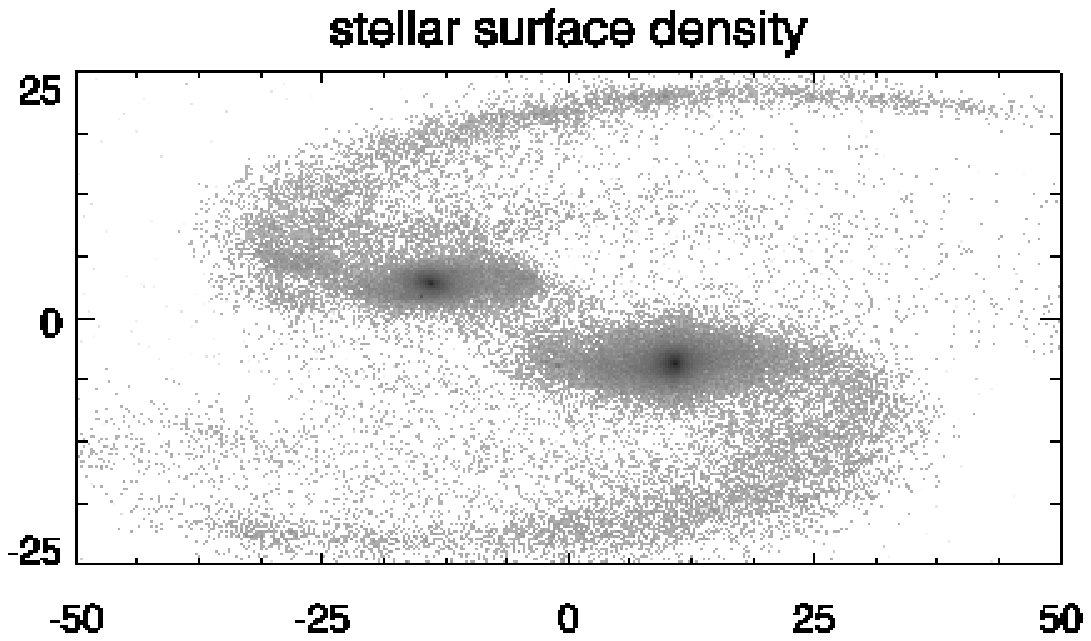}\\
\includegraphics[width=0.49\hsize]{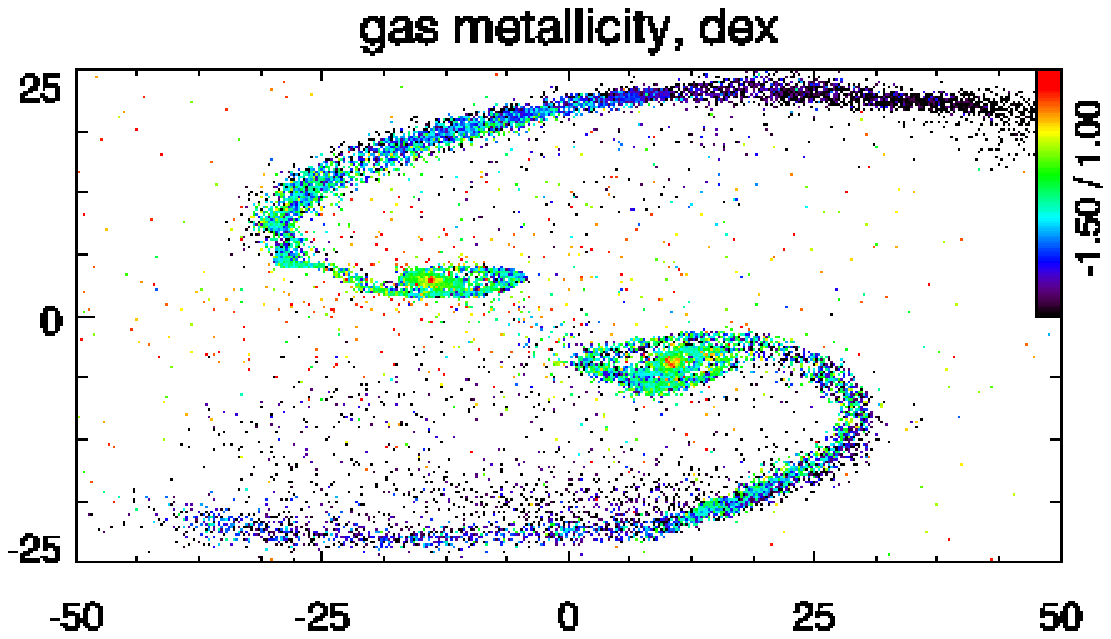}
\includegraphics[width=0.49\hsize]{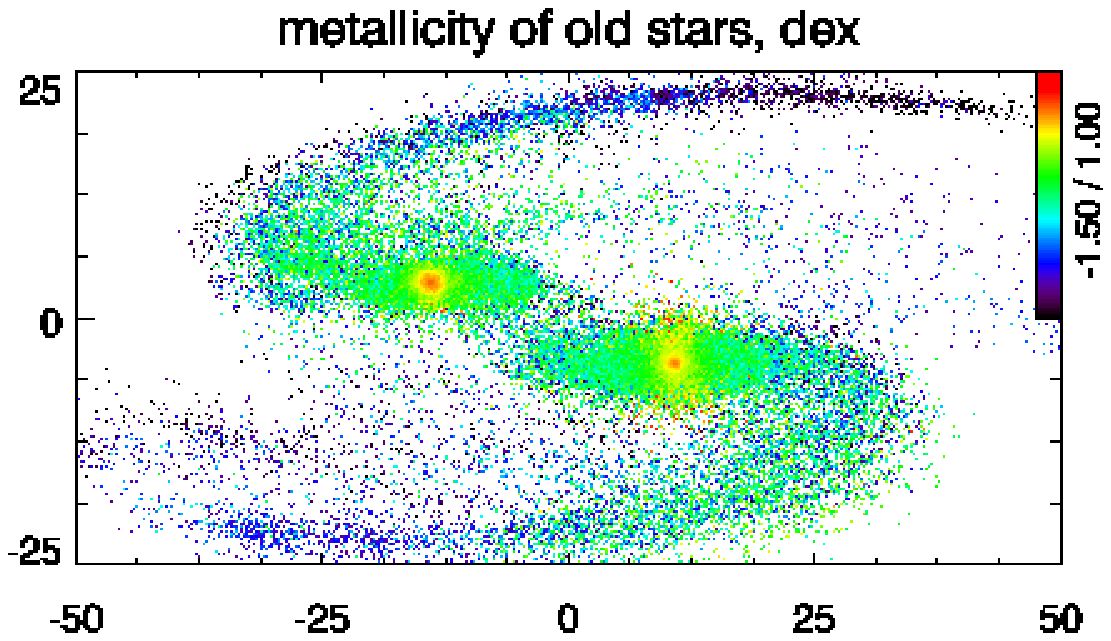}\\
\includegraphics[width=0.49\hsize]{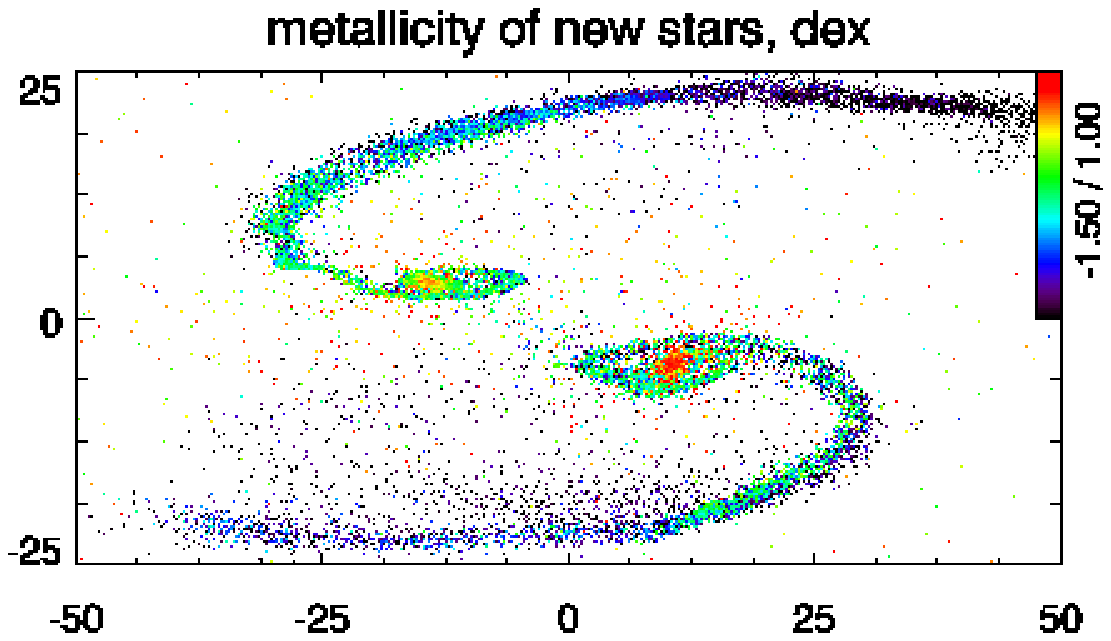}
\includegraphics[width=0.49\hsize]{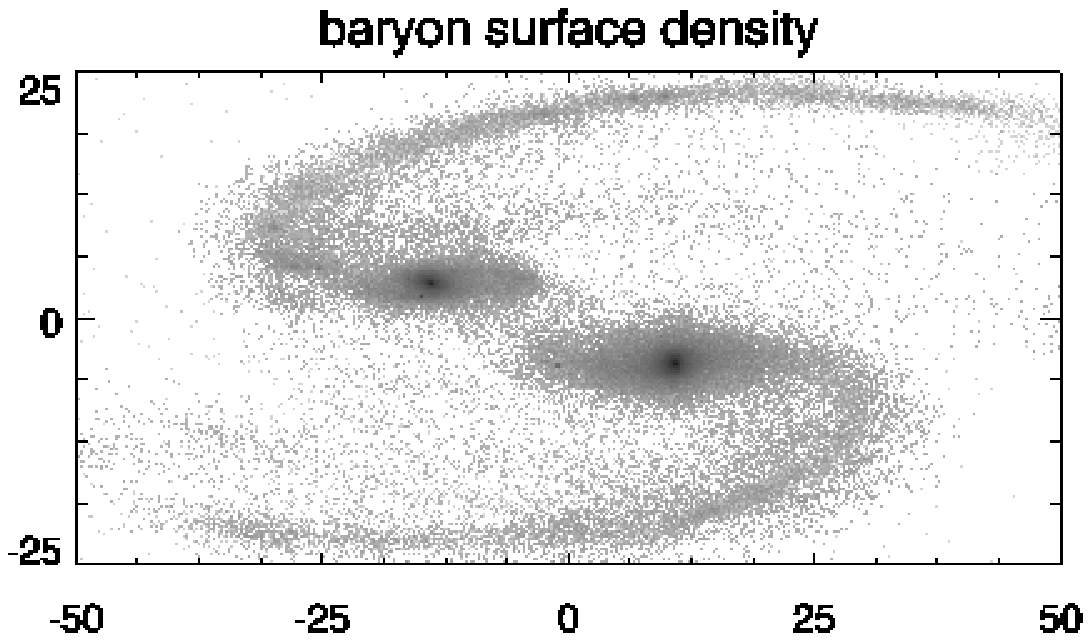}
\caption{Projected maps of 8 quantities traced by the GalMer simulations for 
a merger of gSa and gSb, prograde orbit of type 5, 650~Myr after the start
of the simulations. Left-to-right, top-to-bottom: radial velocity, velocity
dispersion (\emph{star} and \emph{hybrid} particles), projected
density of gas, projected density of old stars, metallicity of gas,
metallicity of old stars, metallicity of new stars, total projected density
(\emph{star} + \emph{hybrid} particles).\label{figMaps}}
\end{figure}

The algorithm is sufficiently fast to perform the map computation in
real time, e.g. the generation of a $400\times400$ pixels map takes a fraction
of a second. In Fig.~\ref{figMaps} we show an example of projected maps for
one of the interactions.

\subsection{Spectrophotometric Properties}

We developed a technique using \soft{PEGASE.HR} \citep{LeBorgne+04}
and \soft{PEGASE.2} \citep{FR97} to model spectrophotometric properties of
interacting galaxies from the results of GalMer simulations. By taking into
account kinematics, star formation (SFH) and chemical enrichment history
(CEH) we model spectra and broadband photometric colours. We will present
all the details regarding the spectrophotometric modelling in a separate
paper (Chilingarian et al. in prep.), but here we will briefly introduce the
algorithms and results which can be obtained.

There were several successful attempts \citep{CW09,JGC10} of modelling
the spectrophotometric properties of the results of $N$-body simulations.
However, in all the known cases the stellar population information is taken
into account in an approximate way, by characterising $N$-body particles by
their mean ages and metallicities. At the same time, the behaviour of
spectral features is a strongly non-linear function of age and metallicity
and it also differs significantly along the wavelength domain, in
particular, due to the dust attenuation. Hence, SEDs computed using
mass-weighted average ages and metallicities may not reflect real spectral
energy distribution in galaxies. With the GalMer simulations, thanks to the
usage of \emph{hybrid} particles we are (1) able to trace SFH and CEH in
detail through the entire duration of the simulation for every particle of
this type; and (2) make use of this information. Therefore, we are able to
make qualitatively better modelling of the spectrophotometric properties of
interacting galaxies (and results of other $N$-body TreeSPH simulations as
well) and make direct comparison with observations. The SFH and CEH are
computed using a whole sequence of snapshots in a given GalMer experiment as
differences of gas masses in each \emph{hybrid} particle in each of the 10
metallicity bins from [Fe/H]$=-2.5$ to $+1.0$~dex.

The most time-consuming part of the modelling is running the
\soft{PEGASE.HR} code. Due to the complexity of the evolutionary synthesis
even with the present state of the art computer hardware it takes several
seconds per spectrum (i.e. per \emph{star} particle), resulting in several
hours per snapshot including up-to 160000 \emph{hybrid} and \emph{star}
particles. Computation of spectra for individual particles is absolutely
crucial to properly model effects of dust extinction and intrinsic
broadening of spectral lines caused by motions of particles along a line of
sight (i.e. internal kinematics of galaxies). However, we can avoid the
actual execution of the evolutionary synthesis code.

The algorithms of evolutionary synthesis such as \soft{PEGASE.HR} include:
\begin{enumerate}
\item computation of isochrones for different ages and metallicities based on
a given set of stellar evolutionary tracks and stellar IMF;
\item picking up stellar spectra from a stellar library (either
empirical or theoretical) for the atmosphere parameters corresponding to a
given point on the isochrone; 
\item co-adding contributions of different types
of stars according to the weights on the isochrone, making up simple stellar
population (SSP) spectrum including stars of a single age and metallicity; 
\item co-adding different SSPs to reproduce complex SFH and CEH.
\end{enumerate}

For a case when the IMF is fixed and SFH and CEH are traced on a pre-defined grid of ages and metallicities, it
becomes possible to execute only the 4th step from the list above. This
means that if we pre-compute only once a grid of SSPs corresponding to a
given IMF (\citet{MS79} in our case) and the grid of ages and
metallicities, we can reuse it for all the particles, improving the
efficiency of the computations \emph{by several orders of magnitude}. Once
for every GalMer experiment we also pre-compute the SFH and CEH and
store them as a two-dimensional histogram for every particle.

For every spatial bin (see definition in the previous section) we first sort
the particles along a line of sight to be able to account for dust
extinction. Then, either a high-resolution spectrum or a broad-band spectral
energy distribution (depending on the mode of the computation) is computed
for every particle starting from the most distant one from the observer.

This is done by co-adding the pre-computed SSPs from the grid mentioned
above with the weights corresponding to a mass contribution of stars of each
age and metallicity contained in the SFH and CEH.

The dust extinction is then applied to the total spectrum or multi-colour
spectral energy distribution (SED) as it was computed at the previous step
(i.e. excluding the current particle). For a known gas column density
and a solar metallicity, we assume a standard dust-to-gas mass ratio
to compute extinction as $A_V = N_H / 1.871 \times 10^{21}$ \citep{BSD78},
where $N_H$ is a number of hydrogen atoms per cm$^2$, assuming $R_V \equiv A_V
/ E(B-V) = 3.1$. We compute $N_H$ along the line of sight and scale this
formula linearly with metallicity. Then, we account for the wavelength
dependence of extinction according to \citet{Fitzpatrick99}, and apply it to
the total stellar population SED generated at this step.

After having applied the dust extinction, the total spectrum of a current
particle is blue- or redshifted according to its radial velocity, which is
done as a simple shift operation with linear interpolation, since our SSP
grid is rebinned with a logarithmic step in the wavelength corresponding to
a fixed pixel size in km~s$^{-1}$.

At the end, the total spectrum (or SED) of a current particle is co-added
to the result.

This algorithm makes it possible to compute in a few seconds for a given
snapshot a total intermediate-resolution ($R=3000$, $3900 < \lambda <
6800$\AA) spectrum, based on the ELODIE.3.1 \citep{PSKLB07} empirical
stellar library. The total number of co-added SSP in this extreme case
reaches $10^6$. The broad-band FUV to NIR SED based on the low-resolution
theoretical BaSeL \citep{LCB97} stellar library is computed much faster,
because the number of SED points in it is only about a dozen compared to
several thousands, allowing us to compute a $400\times400$ pixels
multi-colour datacube in a few seconds.

Due to the limited spatial resolution of GalMer simulations, we are unable
to reach the spatial scales sufficient to take into account properly the gas
clumpiness, therefore we adopt a scale of 250~pc for the computation of
extinction, which roughly corresponds to the resolution of the simulations,
and which results in the values of total extinction in simulated GalMer
isolated galaxies well resembling observations. Our limited resolution also
causes overestimation of extinction effects in central regions with
high gas densities, where large amount of gas falls on-to during the
interaction. Presently, we do not include any dust emission in mid- and
far-IR spectral bands. This will be done in a future release of the GalMer
database and will be presented in a separate paper (Melchior et al. in
prep.)

\begin{figure}
\includegraphics[width=\hsize]{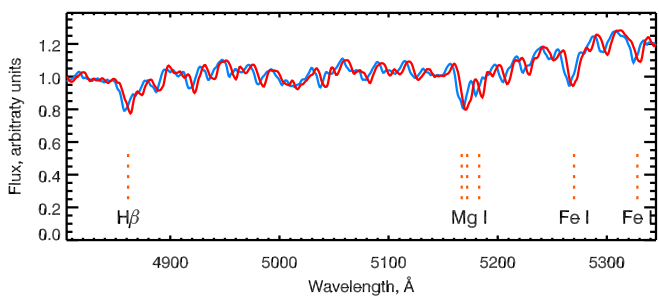}
\caption{Examples of synthetic spectra generated by the spectrophotometric
modelling algorithm applied to the results of GalMer simulations. Two curves
correspond to two different positions along the projected major axis of a
merger remnant: centre approaching (blue), and receding (red). The spectra
are normalised to unity at $\lambda = 5000$~\AA\ for
clarity.\label{figSpec}} 
\end{figure}

In Fig.~\ref{figSpec}, we demonstrate an example of generated spectra for a
merger remnant, where differences caused by the internal kinematics are
clearly visible. Red and blue spectra taken 1.5~kpc from the galaxy centre
along the projected major axis, demonstrate the Doppler shift of spectral
lines. Absorption lines are thus broadened naturally and reflect the
intrinsic stellar velocity dispersion of the galaxy.

At present, we do not attempt to model nebular emission lines, because our
simulations do not trace at sufficient level of detail the physical
conditions in the ISM. Only qualitative modelling can in principle be
performed, assuming fixed ISM temperature and density depending only on the
gas metallicity which is available for \emph{hybrid} particles. This
modelling will be addressed in detail in a separate paper (Melchior et al.
in prep.)

\section{Astrophysical applications}

\subsection{Galaxy properties from modelling}

At present, we explored the GalMer database to study a variety of
physical processes related to galaxy interactions, such as induced star
formation enhancements, evolution of metallicity gradients, angular momentum
redistribution, and its impact on the final kinematical properties of the
merger remnant.

In  \citet{dMCMS07} and  \citet{DiMatteo+08b} we investigated the
enhancement of star formation efficiency in galaxy interactions and mergers.
We showed in \citet{dMCMS07} that, in the final merging phase, retrograde
encounters have greater star formation efficiency than direct encounters,
that the amount of gas available in the galaxy is not the main parameter
governing the star formation efficiency in the burst phase, and that there
is a negative correlation between the amplitude of the star forming burst at
the merging phase and the tidal forces exerted at pericentral passage.  The
general result presented in \citet{DiMatteo+08b} shows that, at low
redshift, galaxy interactions and mergers, in general, trigger only moderate
star formation enhancements. Strong starbursts where the star formation rate
is increased by a factor greater than 5 are rare and found only in about
15$\%$ of major galaxy interactions and mergers. Merger-driven starbursts
are also rather short-lived, with a typical duration of activity of a few
$10^8$ yr. These conclusions are found to be robust, independent of the
numerical techniques and star formation models. At higher redshifts, where
galaxies are gas-rich, gas inflow induced starbursts are neither
stronger nor longer than their local counterparts.  These results are in
good agreement with a number of observational works
\citep{BLA03,Li+08,Jogee+09,KJ09}, demonstrating that interactions
and mergers do not always trigger strong bursts of star formation.

More recently, \citep{DiMatteo+09a},  we investigated  how the
metallicity gradients in dry merger remnants depend on the structure and
metallicity gradients of the galaxies involved in the merger.  Our aim was
to understand if dry mergers could lead to metallicity gradients as observed
in elliptical galaxies in the local Universe, and if they always lead to a
flattening of the initial (i.e., pre-merger) gradient. The analysis of the
whole set of dry merger simulations in the GalMer database allowed us to
show that the ratio of the remnant and the initial galaxy slopes spans a
wide range of values, up to $>1$ (with values greater than one resulting
only when companions have gradients twice that of the progenitor).
For a merger between two ellipticals having identical initial metallicity
slopes (i.e., equal companion and galaxy slopes), the metallicity profile of
the remnant flattens, with a final gradient about 0.6 times the initial one.
This flattening depends neither on the characteristics of the orbit of the
progenitors or on their initial concentration. If the companion slope is
sufficiently steep, ellipticals can maintain their original pre-merger
metallicity gradient. These results, compared to the observed variety of
metallicity gradients in dwarf elliptical and lenticular galaxies
\citep{Chilingarian09}, may suggest the mergers to be an important channel
of dE/dS0 evolution.

Given the diversity in outcomes of the mergers, we concluded that dry
mergers do not violate any observational constraints on the systematic
characteristics of metallicity gradients in local ellipticals
\citep{ogando+05}.

The redistribution of the orbital angular momentum into internal rotation,
and its impact on the kinematical properties of the merger remnant,
became a subject of two studies \citep{dMCMS08,DiMatteo+09b}.

In \citet{dMCMS08}, we presented a new scenario to form
counter-rotating central components in early-type galaxies, by dissipative
and dissipationless ``mixed'' mergers, consisting of elliptical-spiral
systems in retrograde orbits. We demonstrated that the counter-rotation
can appear both in dissipative and dissipationless retrograde mergers, and
it is mostly associated to the presence of a disc component, which preserves
part of its initial spin. In turn, the external regions of the two
interacting galaxies acquire part of the orbital angular momentum, due to
the action of tidal forces exerted on each galaxy by the companion.  In the
case of dissipative mergers, the central decoupled core could be composed of
two distinct populations: the old stellar population, which has preserved
part of its initial spin, and a new stellar population, born \emph{in situ}
from the kinematically decoupled gas component.

Even the merger of two initially non-rotating, pressure supported progenitor
galaxies can lead to remnant galaxies having peculiar kinematical
properties. In \citet{DiMatteo+09b}, we have indeed shown that it is
possible to generate elliptical-like galaxies, with $v/\sigma >$ 1 outside
one effective radius, as a result of the conversion of orbital into
internal angular momentum. This conversion occurs ``outside-in'': the
external regions acquiring part of the angular momentum first, and it affects
both the baryonic and the dark matter components of the remnant galaxy (i.e.
both acquire part of the angular momentum, the relative fractions depend on
the initial concentration of the merging). If the initial baryonic component
is sufficiently dense and/or the encounter takes place on a orbit with high
angular momentum, the remnant galaxy exhibits hybrid properties, i.e. an
elliptical-like morphology, but rotational support in the outer stellar halo
($v/\sigma >$ 1). Systems with these properties have been recently observed
through a combination of stellar absorption lines and planetary nebulae for
kinematic studies of early-type galaxies \citep{Coccato+09}. Our results are
in qualitative agreement with such observations and demonstrate that even
mergers composed of non-rotating, pressure-supported progenitor galaxies can
produce early-type galaxies with significant rotation at large radii.
 
\subsection{Synthetic observations: virtual telescope}

Some of the value-added data analysis services we provide for the GalMer
database can be used as a ``virtual telescope'' to create simulated images
and spectra of interacting galaxies. Thanks to the high-quality
stellar population modelling provided by the \soft{PEGASE.2/PEGASE.HR} code,
we are able to compare the simulated data to real observations.

\begin{figure}
\includegraphics[width=\hsize]{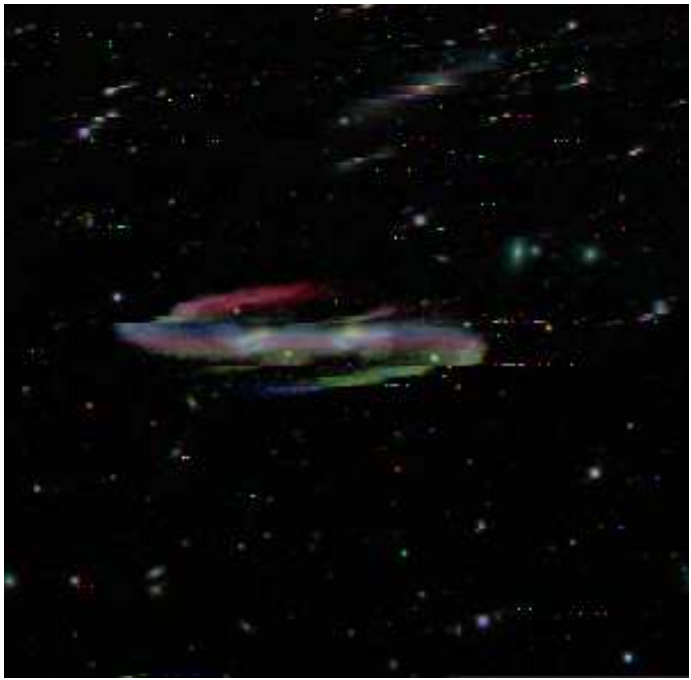}
\caption{RGB false-colour image of a simulated galaxy merger displayed over
the SDSS image of the Abell~85 galaxy cluster. \label{figRGB}}
\end{figure}

In Fig.~\ref{figRGB}, we present an RGB false-colour composite image of an
interacting galaxy pair created from the results of spectrophotometric
modelling described above applied to GalMer simulations. The
simulation result is superimposed over the background showing the
Sloan Digital Sky Survey Data Release 7 \citep{SDSS_DR7} image of the galaxy
cluster Abell~85. We used SDSS $g$, $r$, and $i$ bands in both observational
and simulated data, and the RGB visualisation code implementing the
algorithm by \citet{Lupton+04} to generate a false-colour image. Since
the results of our spectrophotometric modelling are expressed in physical
units, they can be transformed into observables (mag~arcsec$^{-2}$) and
projected on-to the sky (see eq.\ref{xyproj}), providing a direct way of
comparing them to observational data. By co-adding model broadband fluxes
and those coming from SDSS direct imaging, we can create realistically
looking false-colour images such as the one shown in Fig.~\ref{figRGB}.

Simulated images can then be analysed using classical observational
techniques, e.g. surface photometry. A successful example of such analysis
is demonstrated in \citet{Chilingarian+09}, where we found a match
between a rather complex three-component observed density profile of the
lenticular galaxy NGC~6340 and a merger remnant from the GalMer simulations,
supporting the major merger to be an important event in the evolution of
this particular object.

We are also able to simulate the whole data cubes corresponding to the data
produced by modern integral-field unit (IFU) spectrographs, such as SAURON
at 4.2~m William Herschel Telescope, VIMOS at ESO Very Large Telescope, and
GMOS at Gemini. Our spectral resolution and coverage allows us to model
high-resolution blue setting of VIMOS (HR-Blue, $R=2500$) and B600 grating
of GMOS ($R=2200$) usually chosen to study stellar kinematics of nearby
galaxies.

The spatial resolution of our simulations (0.20$-$0.28~kpc) is comparable to
that of the SAURON \citep{deZeeuw+02} survey targeting nearby galaxies at
distances between 10 and 40~Mpc, thus, having spatial resolution of 0.05 to
0.2~kpc given the average atmosphere seeing quality of 1~arcsec. At the same
time, we can degrade the spatial resolution of our simulation in order to
match observations made with VIMOS, e.g. post-starburst galaxies presented
in \citet{CdRB09}.

As the stellar population models improve, we will be able to upgrade the
spectrophotometric modelling engine in the GalMer database in order to
simulate data coming from next generation IFU facilities such as the
second-generation VLT instrument MUSE.

\subsection{Colour-magnitude relations}

The spectrophotometric modelling facilities allow us to compute total
magnitudes and colour of interacting galaxies and to directly compare them
to observational data.

\begin{figure}
\includegraphics[width=\hsize]{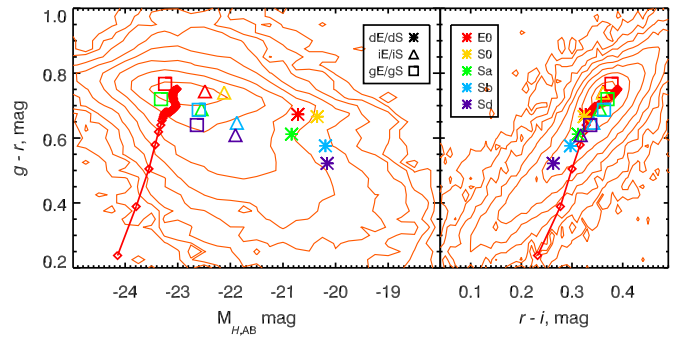}
\caption{Colour--magnitude (left) and colour-colour relations demonstrating
contour density plots for 80000 observed low-redshift ($z<0.3$) SDSS
galaxies, GalMer galaxy models corresponding to isolated galaxies of
different masses and morphological types, and a sequence of a GalMer 
simulation of a giant--giant interaction followed from the merger epoch for
1.5~Gyr every 50~Myr (red diamonds). 
Colours (see right panel legend) code morphological type, while symbol shapes 
(asterisk, triangle and square, see left panel legend) code luminosity-class.
\label{figColMag}}
\end{figure}

The colour--magnitude and colour--colour plots comparing results of GalMer
simulations to a large observational dataset are presented in
Fig.~\ref{figColMag}. A well-known galaxy colour bimodality \citep[see
e.g.][]{Strateva+01} is clearly seen in the contour plot presenting the
distribution of some 80000 nearby galaxies ($z<0.3$) from the SDSS survey. The
photometric data were corrected for the Galactic extinction according
to \citet{SFD98} and converted into rest-frame, i.e. $k$-corrected
using the analytical approximations presented in Chilingarian et al.
(submitted). The ``red sequence'' and the ``blue cloud'' separated by the
``green valley'', a locus of post-starburst galaxies \citep[e.g.][]{Goto+03}
can be identified.

The total magnitudes of GalMer mergers in SDSS bands are computed using the
spectrophotometric modelling described above. We show the positions of
GalMer galaxy models (isolated galaxies) of different masses and
morphological types. Elliptical and lenticular galaxies (red and yellow
symbols) are sitting on the red sequence, while spiral galaxies are bluer,
directed down toward the blue cloud. GalMer Sa and Sb spiral galaxies
(green and blue symbols) at an evolutionary stage shown in
Fig.~\ref{figColMag} reside in the green valley.

We also follow the evolution of an interacting pair from the time when
galaxies merged (red diamonds connected with lines) till the end of the
simulation showing intermediate snapshots every 50~Myr. Right after the
gas-rich merger, remnants sit in the ``blue cloud'', despite very strong
internal extinction of regions of massive star formation. Then, as the star
formation decreases, luminosity-weighted ages of stellar populations
increase causing global colours to become redder and, thus, moving merger
remnants in the top-right direction across the ``blue cloud'' through the
``green valley'' toward the ``red sequence''. Finally, 400--600~Myr after
the merger time, all remnants settle on-to the ``red sequence'', although
some residual star formation is still observed.

One can see that merger remnants cross the ``green valley'' very fast giving
possible explanation why there is at present a deficit of galaxies in this
region of the colour--magnitude diagram, given that all mergers in this mass
range have not yet been included in the database.

\section{Summary}

We present the GalMer database providing online access to the results of
TreeSPH simulations. The structure of the database conforms to recent
International Virtual Observatory standards. We describe the interactive
data access web-interface, advanced mechanisms for data visualisation and
manipulation using connection to the Virtual Observatory tools dedicated for
dealing with tabular, imaging and spectral data. 

After having stored the snapshots in FITS binary table format, giving access
to all coordinates and data on individual simulation particles,
we provide a set of value-added tools using the results of TreeSPH
simulations. These include: (1) generator of projected maps of various
physical quantities traced by the simulations; (2) engine to perform the
modelling of spectrophotometric properties of interacting and merging
galaxies based on the \soft{PEGASE.2/PEGASE.HR} evolutionary synthesis code.
The latter tool can be used as a virtual telescope and synthetic images,
spectra, and datacubes generated using it are directly comparable to
observational data.

We provide examples of several use-cases for the GalMer database using our
value-added tools.

The database will be updated by including new simulations as they have been
completed.

\begin{acknowledgements}

This research used computational resources of the Informatic Division of the
Paris Observatory, Commissariat \`a l'\'Energie Atomique,
the CNRS national centre IDRIS, and those available
within the framework of the Horizon project. A server
dedicated to the on-the-fly computations and the database in Paris
Observatory has been provided by the VO Paris Data Centre. IC acknowledges
additional support from the RFBR grant 07-02-00229-a. 

Funding for the SDSS and SDSS-II has been provided by the Alfred P. Sloan
Foundation, the Participating Institutions, the National Science Foundation,
the U.S. Department of Energy, the National Aeronautics and Space
Administration, the Japanese Monbukagakusho, the Max Planck Society, and the
Higher Education Funding Council for England. The SDSS Web Site is
http://www.sdss.org/.

\end{acknowledgements}

\bibliographystyle{aa}
\bibliography{galmer}

\begin{appendix}
\section{Galaxy models: density profiles}\label{profiles}

In Sect.\ref{galmod}, we presented the galaxy models adopted for the
simulations. In particular, we saw that the dark halo and the optional
bulge are modelled as a Plummer sphere \citep[][pag.42]{BT87}, with
characteristic masses $M_B$ and $M_H$ and characteristic radii $r_B$ and
$r_H$. Their densities are given, respectively, by the following analytical
formula:

\begin{equation}\label{halo}
\rho_{H}(r)=\left(\frac{3M_{H}}{4\pi {r_{H}}^3}\right)\left(1+\frac{r^2}{{r_{H}}^2}\right)^{-5/2}
\end{equation}
and
\begin{equation}\label{bulge}
\rho_{B}(r)=\left(\frac{3M_{B}}{4\pi {r_{B}}^3}\right)\left(1+\frac{r^2}{{r_{B}}^2}\right)^{-5/2}.
\end{equation}
The stellar and gaseous discs follow a Miyamoto-Nagai density profile
\citep[][pag.44]{BT87}:

\begin{eqnarray}\label{stdisc}
\rho_{*}(R,z)&=&\left(\frac{{h_{*}}^2 M_{*}}{4 \pi}\right)\times\nonumber\\&&\frac{a_{*} R^2+(a_{*}+3\sqrt{z^2+{h_{*}}^2})\left(a_{*}+\sqrt{z^2+{h_{*}}^2}\right)^2}
{\left[a_{*}^2+\left(a_{*}+\sqrt{z^2+{h_{*}}^2}\right)^2\right]^{5/2}\left(z^2+{h_*}^2\right)^{3/2}}
\end{eqnarray}
\begin{eqnarray}\label{gasdisc}
\rho_{g}(R,z)&=&\left(\frac{{h_{g}}^2 M_{g}}{4 \pi}\right)\times\nonumber\\&&\frac{a_{g} R^2+(a_{g}+3\sqrt{z^2+{h_{g}}^2})\left(a_{g}+\sqrt{z^2+{h_{g}}^2}\right)^2}
{\left[a_{g}^2+\left(a_{g}+\sqrt{z^2+{h_{g}}^2}\right)^2\right]^{5/2}\left(z^2+{h_g}^2\right)^{3/2}},
\end{eqnarray}
with masses $M_{*}$ and  $M_{g}$ and  vertical and radial scale lengths given, respectively, by $h_{*}$ and $a_{*}$, and $h_{g}$ and $a_{g}$.\\

\section{Galaxy models evolved in isolation}\label{isolated}
 
\begin{figure}
\includegraphics[width=\hsize]{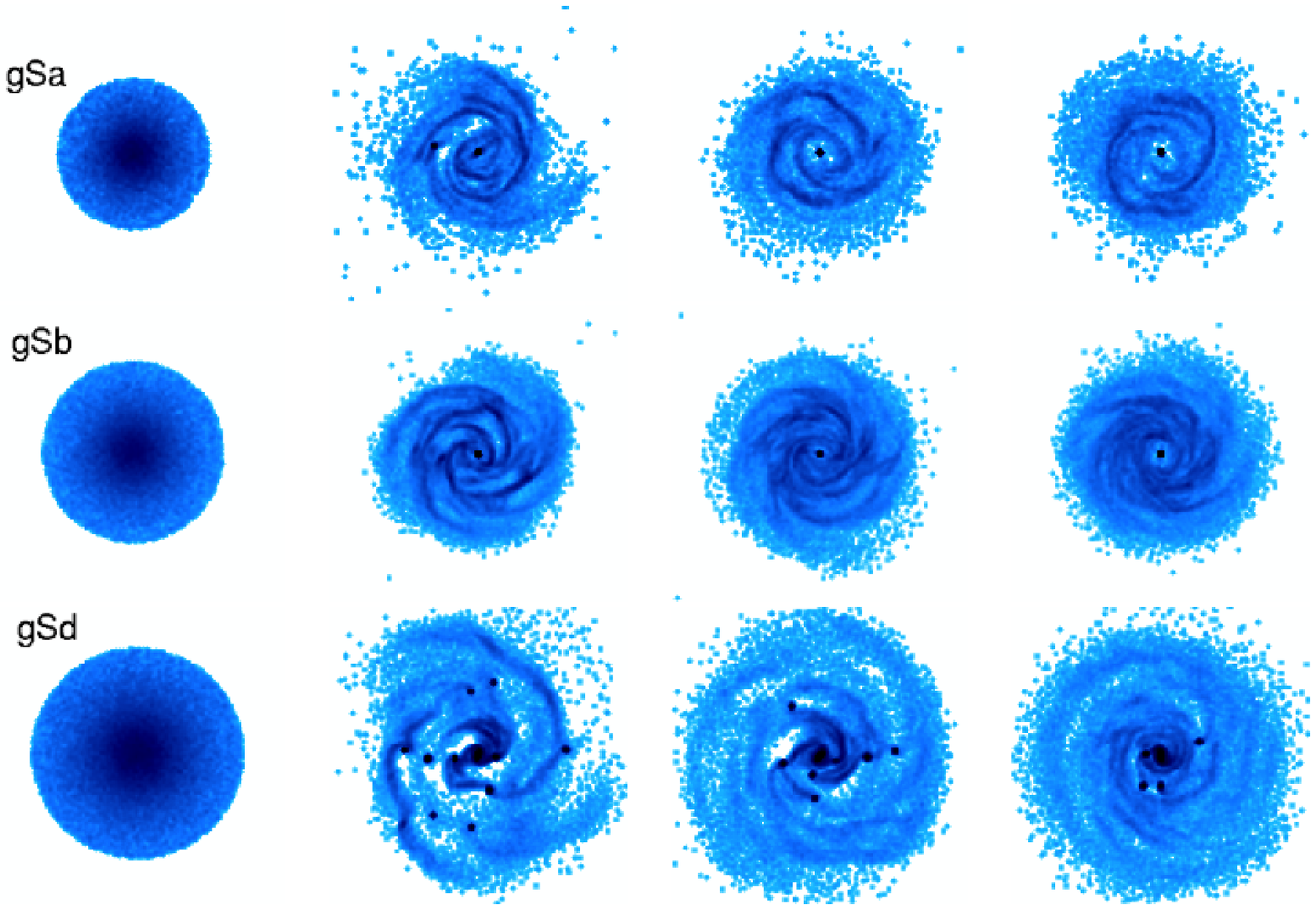}
\caption{Gas maps of the gSa, gSb, and gSd models with $Q_{\mathrm{gas}}=0.3$ evolved in isolation. Snapshots are equally spaced in time from t=0 to t=3 Gyr. Each map is 60 kpc x 60 kpc in size. }\label{gas_iso}
\end{figure}

\begin{figure}
\includegraphics[width=\hsize]{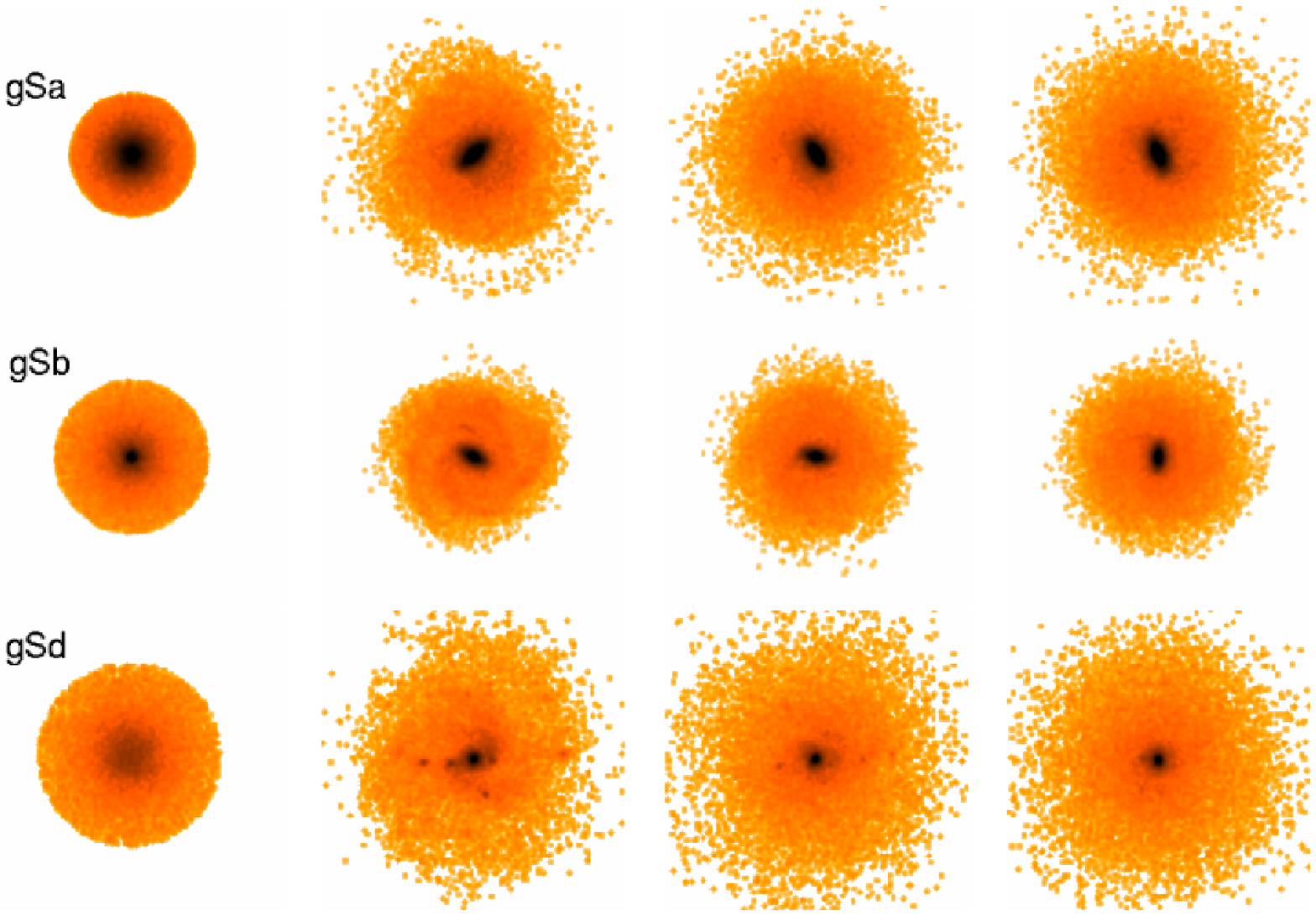}
\caption{Star maps of the gSa, gSb, and gSd models evolved in isolation. Snapshots are equally spaced in time from t=0 to t=3 Gyr. Each map is 60 kpc x 60 kpc in size. }\label{star_iso}
\end{figure}

Since the aim of this work is to study the effects that  interactions and
mergers have in driving galaxy evolution, it is important to study, for
comparison, also the evolution of the galaxies in isolation, in order to
distinguish properly secular processes from those related to the
interaction. In Figs. \ref{gas_iso}, and \ref{star_iso}, we show the
evolution of the gaseous and stellar components of the giant gSa, gSb and
gSd galaxies, respectively, evolved in isolation for 3 Gyr. As it can be
seen, all the models evolve rapidly in the first 0.5-1 Gyr: a stellar bar
and spiral arms form, and at the same time gas compresses into density waves
and clumps and partially falls into the central regions, where a star
formation enhancement takes place (for the star formation rate of these
galaxies, we refer the reader to \citealp{dMCMS07}).


\end{appendix}

\end{document}